\documentclass[useAMS,usenatbib]{mnras}

\usepackage[utf8]{inputenc}
\usepackage{graphicx}
\usepackage{color}
\usepackage{booktabs}
\usepackage{amssymb}
\usepackage[]{natbib}

\newcommand{\eagle}{{\sc eagle}}
\newcommand{\gama}{{\sc gama}}
\newcommand{\simref}{{\sc eagle}}     
\newcommand{\simdmo}{{\sc eagle-dmo}} 
\newcommand{\pmil}{{\sc p-millennium}}
\newcommand{\gadget}{{\sc gadget}}
\newcommand{\galform}{{\sc galform}}
\newcommand{\illustris}{{\sc illustris}}

\def\apj{ApJ}
\def\apjs{ApJS}
\def\apjl{ApJ}
\def\aj{AJ}
\def\mnras{MNRAS}

\def\aap{A\&A}
\def\pasp{PASP}

\defcitealias{Farrow15}{F15}

\title[Galaxy clustering in \eagle]{Small-scale galaxy clustering in the \eagle\ simulation}
\author[M. C. Artale, et al.]{M. Celeste Artale$^1$\thanks{E-mail:mcartale@iafe.uba.ar}, Susana E. Pedrosa$^1$, James W. Trayford$^2$, Tom Theuns$^2$, \newauthor Daniel J. Farrow$^4$,  Peder Norberg$^{2,3}$, Idit Zehavi$^5$, Richard G. Bower$^{2,3}$,\newauthor and Matthieu Schaller$^2$\\
$^1$ Instituto de Astronom\'ia y F\'isica del Espacio (IAFE,CONICET-UBA), C.C. 67 Suc. 28, C1428ZAA Ciudad de Buenos Aires, Argentina\\
$^2$ Institute for Computational Cosmology, Department of Physics, Durham University, South Road, DH1 3LE Durham, UK\\
$^3$ Centre for Extragalactic Astronomy, Department of Physics, Durham University, South Road, DH1 3LE Durham, UK \\
$^4$ Max-Planck-Institut f\"ur extraterrestrische Physik, Postfach 1312 Giessenbachstrasse, D-85741 Garching, Germany\\
$^5$ Department of Astronomy, Case Western Reserve University, 10900 Euclid Avenue, Cleveland, OH 44106, USA\\
} 
 
\begin{document}
\maketitle

\begin{abstract}
 We study present-day galaxy clustering in the \eagle\ cosmological hydrodynamical simulation. \eagle's galaxy formation parameters were calibrated to reproduce the redshift $z=0.1$ galaxy stellar mass function, and the simulation also reproduces galaxy colours well. The simulation volume is too small to correctly sample large-scale fluctuations and we therefore concentrate on scales smaller than a few mega parsecs. We find very good agreement with observed clustering measurements from the Galaxy And Mass Assembly (\gama) survey, when galaxies are binned by stellar mass, colour, or luminosity. However, low-mass red-galaxies are clustered too strongly, which is at least partly due to limited numerical resolution. Apart from this limitation, we conclude that \eagle\ galaxies inhabit similar dark matter haloes as observed \gama\ galaxies, and that the radial distribution of satellite galaxies as function of stellar mass and colour is similar to that observed as well.
\end{abstract}


\section{Introduction}
The spatial distribution of galaxies provides a powerful way to probe both cosmology and galaxy formation. Galaxy clustering measurements on scales where density fluctuations are only mildly non-linear, combined with other cosmological data sets such as CMB measurements, put impressively tight constraints on cosmological parameters \citep[e.g.][]{Hinshaw13, Planck16}. In addition, the detection of baryon acoustic oscillations in the clustering of galaxies \cite[e.g.][]{Cole05, Eisenstein05} opened up the way to quantify the nature of dark energy \citep[e.g.][]{Euclid11}, and combined with redshift-space distortion measurements, test theories of gravity \citep[e.g.][]{Linder08}. 

From the perspective of galaxy formation, the clustering of galaxies inform us about the relation between galaxies and the underlying dark matter and can provide hints about the physical processes involved in galaxy assembly history. As galaxies reside within the dark matter haloes, their positions trace the underlying cosmic structure. While the formation and evolution of the dark matter haloes is governed exclusively by gravitational interaction, the assembly of the galaxies is governed by the more
complex baryon physics that also affects the distribution of galaxies. 
Such \lq galaxy bias\rq\ may impact as well cosmological inferences made from galaxy clustering measurements. 

The main statistical tool used for characterising galaxy clustering is the two-point correlation function $\xi$(r), which measures the excess probability over random of finding pairs of galaxies at different separations $r$ \citep{Peebles1980}. Commonly, when analysing redshift surveys, the projected correlation function integrated along the line-of-sight is used, in order to eliminate in principle
redshift-space distortions \citep{Davis1983}. 
Observations show that brighter, redder and more massive galaxies are more strongly clustered and related trends are also measured as a function of morphology and spectral type (e.g.,
\citealt{Norberg2002,Zehavi2002,Zehavi2005,Goto2003,Li2006,Coil2006,Croton2007b,Zheng2007,Coil2008,Zehavi2011,Coupon2012,Guo2013,Farrow15}). 

The theoretical modelling of galaxies plays an important role in interpreting clustering data, since, although galaxy clustering on large enough scales is very similar to that of the underlying matter distribution, it is not expected to be identical. Such models are also routinely used to estimate sample variance and verify methods for correcting observational biases. Several theoretical schemes are able to model galaxy clustering in volumes comparable to those probed observationally. These start from a dark matter only (DMO) $N$-body simulation, and populate haloes or sub-haloes with galaxies. Halo occupation distribution models 
\citep[{\sc hod}, e.g.][]{Cooray02,Berlind02,Tinker12} or sub-halo abundance matching 
\citep[{\sc sham}, e.g.][]{Conroy06,Vale04,Vale06} are statistical techniques that match galaxies with haloes by abundance based on, for example, their circular velocity. Semi-analytical galaxy formation techniques \citep[e.g.][]{Kauffmann93,Cole94} use physically motivated schemes to associate galaxies to haloes \citep[see e.g.][for a review]{Baugh06}.

Notwithstanding the successes of these methods, they suffer from intrinsic limitations. Stellar and AGN feedback from forming galaxies affect the mass of their halo \citep{Sawala13,Velliscig2014,Schaller15}, limiting the extent to which any DMO simulation predicts the clustering of haloes as function of mass accurately. Feedback effects are plausibly strong enough to affect the mass distribution itself \citep{vanDaalen2011,Sembolini11} and with it galaxy clustering \citep{Hellwing16}. These effects may be relatively small, but the main limitation of the models is on smaller scales, where several effects that occur when a galaxy becomes a satellite (tidal interactions, ram-pressure stripping, strangulation) come into play.
\citet{vanDaalen2014} investigate how physics behind galaxy formation affect the clustering of 
galaxies at small scales through comparing two models of {\sc owls} project \citep{Schaye10} with and without AGN feedback. They found that the physics of galaxy formation affects clustering on small scales, and in addition affects larger scales through its impact on the masses of sub halos.
\citet{Farrow15} show how clustering in galaxy light cone mocks generated with two versions of the semi-analytical \galform\ code
\citep{Lacey2016,GonzalezPerez2014} differ significantly with the observations on small scales. McCullagh et al. ({\em in prep.}) 
and \citet{Gonzalez2017} show that once a more detailed merger scheme is considered, such as that described by \cite{Simha13} reasonably good agreement with clustering data on small scale is achieved. This is in agreement with the findings of \cite{Contreras2013} in which different families of galaxy formation models were compared to clustering data.
\cite{McCarthy16} present results from the {\sc bahamas} cosmological hydrodynamical simulation. These simulations were designed using a similar calibration strategy as \eagle\ but using the {\sc owls} implementation of galaxy formation described by \cite{Schaye10}. The simulated volume (400~Mpc/$h$ on a side) and mass resolution (initial baryonic particle mass $8\times 10^{8}h^{-1}{\rm M}_\odot$) allow them to probe the galaxy correlation function on large scales, but not to go to lower-mass galaxies and smaller scale clustering that we concentrate on here, nor to investigate clustering as function of galaxy property such as for example colour. \cite{Sales2015} compare the distribution of satellite galaxies from the {\sc illustris} simulation \citep{Vogelsberger2014} as function of colour to Sloan Digital Sky Survey ({\sc sdss}, \citealt{York00}) observations by \cite{Wang14}. They attribute the better agreement of the simulations compared to semi-analytical models to the more realistic gas contents of satellites at infall.
	
Galaxy clustering on small scales, and as a function of intrinsic galaxy properties such as luminosity, colour and star formation rate, thus might prove to be a stringent test of galaxy formation models. Performing such a test is the aim of this paper: we explore the clustering of galaxies in the cosmological hydrodynamical \eagle\ simulation \citep{Schaye15} and its dependence on galaxy properties. The galaxy formation model of \eagle\ uses sub-grid modules that are calibrated to reproduce the present-day stellar mass function \citep[as described by][]{Crain15}. In addition, the \eagle\ simulation reproduces relatively well the colours and luminosities of galaxies both in the infra-red \citep{Camps16} and at optical wavelengths \citep{Trayford16}.

The 100~Mpc extent of the largest \eagle\ simulation volume analysed in this paper is too small to properly sample large-scale modes, and, as is well known, such missing large-scale power quantitatively affects clustering measures even on smaller scales \cite[e.g.][]{Bagla05, Bagla06,Trenti2008}. To estimate the severity of this, we compare the clustering of haloes in a dark matter- only version of the \eagle\ volume to that in a much larger volume, simulated with the same cosmological parameters. This allows us to estimate the limitations of our approach.
	We compare the \eagle\ predictions to clustering measurements by \cite{Farrow15} of galaxies in
	the Galaxy and Mass Assembly redshift survey  \citep[\gama,][]{Driver2011,Liske2015}, which are
	in accord with the \cite{Zehavi2011} {\sc sdss} measurements. For completeness we note that
	\cite{Crain16} show that \eagle\ reproduces the observed clustering of $z=0$ H{\sc i} sources.
 	 	 	
This paper is organised as follows. In ~\S~\ref{sec:sim_data} we describe the main characteristics of the simulations used, and briefly discuss the {\gama} survey to which we compare. In ~\S~\ref{sec:methods} we define the notation and present the tools used to measure galaxy clustering. Simulations and observations are compared in ~\S~\ref{sec:results}. In the discussion section, \S~\ref{sec:discuss}, we compare \eagle\ with the clustering in 
{\sc galform} \citep[following the analysis of][]{Farrow15} and our own clustering measurements using the database of the {\sc illustris} \citep{Vogelsberger2014} simulation. The conclusions are summarised in ~\S~\ref{sect:conclusions}.

Throughout this paper and unless specified otherwise, we use the \cite{Planck13} values of the cosmological parameters ($\Omega_{\rm b} = 0.0482$, $\Omega_{\rm dark} = 0.2588$, $\Omega_\Lambda = 0.693$  and $h = 0.6777$, where $H_0 = 100\; h$ km s$^{-1}$ Mpc$^{-1}$. Observational measures of clustering are (most commonly) specified in \lq $h$\rq-dependent units, and to ease comparison to other clustering studies, we will express distances in $h^{-1}$~Mpc and masses in $h^{-2}$M$_\odot$.

\section{Simulations and Data}
\label{sec:sim_data}

This section briefly describes the simulations used, and the \gama\
survey to which clustering results of the simulations are compared.

\begin{table} 
\begin{center}
\caption{Numerical parameters of cosmological simulations considered. From
left to right: simulation identifier, simulation co-moving side length 
$L$, initial mass $m_{\rm g}$ of baryonic particles, dark 
matter particle mass, Plummer-equivalent co-moving length ($\epsilon_{\rm com}$) 
and maximum proper gravitational softening length $\epsilon_{\rm prop}$. The \eagle\ and \simdmo\ simulations are referred to as L0100N1504 and 
L0100N1504-DMO by \citet{Schaye15}.
} 
\label{tab:sims}
\begin{tabular}{lrrrrr}
\hline
Name & $L$ & $m_{\rm g}$ & $m_{\rm dm}$ & $\epsilon_{\rm com}$ & $\epsilon_{\rm prop}$ \\  
& $h^{-1}$Mpc & $10^6\,{\rm M}_\odot$ & $10^6\,{\rm M}_\odot$ & kpc & kpc\\
\hline 
\eagle\          & 67.77 & 1.81  & 9.70 & 2.66  & 0.7\\
\simdmo\         & 67.77 & -     & 11.51 & 2.66  & 0.7\\
\pmil\    & 542.16 & -     &  157 & 3.40 & 3.40 \\
\hline
\end{tabular}
\end{center}
\end{table}

\subsection{The \eagle\ hydrodynamical simulation suite}
\label{sec:sim}
\begin{figure}
  \centering
  \includegraphics[width=\columnwidth]{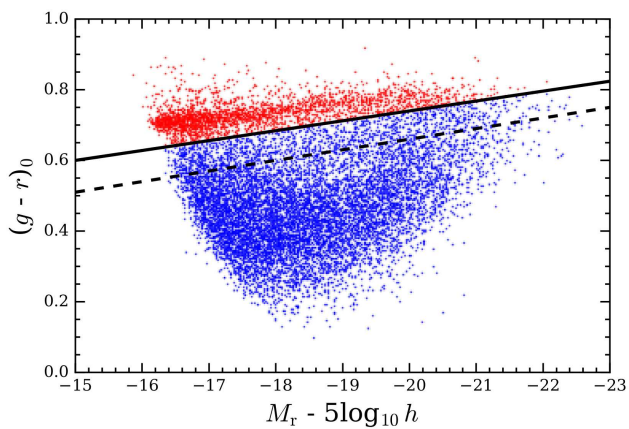}
  \caption{Distribution of rest-frame $(g-r)_{0}$ colour versus $r$-band absolute magnitude, 
    for $z=0.1$ \eagle\ galaxies (coloured points). The rapid colour-dependent decline of galaxies fainter than $M_{\rm r}-5\log_{10}(h)=-17$ results from imposing a stellar mass cut of $M_{*} > 10^{8.66} h^{-2} M_{\odot}$.
    The \textit{solid black line} 
    from Eq.~(\ref{eq:col_split_eagle}) distinguishes red from blue galaxies, \eagle\ galaxies above (below) this line
    line are represented by a red (blue) dot. The \textit{black dashed-line} is the corresponding colour-cut from
    \citet{Farrow15} for \gama\ galaxies (their Eq.~4, and Eq.~\ref{eq:col_split_gama} in the text)}
  \label{fig:col-mag}
\end{figure}

We use the \lq reference\rq~\eagle\ simulation from Table~2 in \cite{Schaye15} 
(i.e.\ L0100N1504, but hereafter referred to as the \eagle\ simulation),
a hydrodynamical cosmological simulation that start at $z=127$ from initial 
conditions generated using the {\sc panphasia} multi-resolution phases of 
\cite{Jenkins13}, taking the \cite{Planck13} cosmological parameter values.
The simulation is part of the \eagle\ simulation suite 
\citep{Schaye15, Crain15} and Table~\ref{tab:sims} lists some of the key simulation
parameters. The \eagle\ simulations were performed with the \gadget-3 code, 
which is based on \gadget-2\ \citep{Springel05}, and uses \lq sub grid models\rq, briefly discussed in more detail below, to encode physical processes below the resolution limit. These models are formulated using parameters or functional forms that express  limitations in our understanding of a given process (for example star formation) or our inability to simulate accurately a known process because of lack of numerical resolution (e.g. the effect of a supernova explosion on the interstellar medium in the presence of radiative cooling). These parameters and functions are calibrated so that the simulation reproduces a limited set of observed properties of galaxies, by performing a large set of simulations in which these parameters are varied as described by \cite{Crain15}. The set of constraints is limited, mainly because each simulation takes a long time to run. In the case of \eagle, sub grid parameters were calibrated to observations at $z\approx 0$ of 
the galaxy stellar mass function of \cite{Baldry2012}, galaxy sizes as measured by \cite{Shen2003}, and the relation between black hole mass and stellar mass.

The hydrodynamics used in \eagle\ uses a number of improvements to the SPH 
implementation collectively referred to as {\sc anarchy} and described by 
Dalla Vecchia ({\em in prep.}); see \cite{Schaller15} for a discussion of 
the relatively small impact of these changes on the properties of simulated 
galaxies. We briefly summarise below the sub grid modules for unresolved physics 
relevant for this paper:

\begin{itemize}
\item {\em Photo-heating and radiative cooling} by the optically thin
  evolving UV/X-ray background of \cite{Haardt01} is
  implemented element-by-element as described by \cite{Wiersma09a}. 
\item {\em Star formation} is implemented using the pressure law
  of \cite{Schaye08} and the metallicity-dependent star
  formation threshold of \cite{Schaye04}. Gas particles eligible for
  star formation are converted to \lq star particles\rq\
  stochastically with a probability that depends on their star
  formation rate and their time step.
\item {\em Stellar evolution and enrichment} is implemented as
  described by \cite{Wiersma09b}: assuming stars form with the
  \cite{Chabrier03} stellar initial mass function (IMF), spanning the
  range [0,1,100]~M$_\odot$, we use stellar evolution and yield
  tables to calculate the rate of type~Ia and type~II (core-collapse)
  supernovae, and follow the rate at which stars enrich the
  interstellar medium through AGB, type~Ia and type~II evolutionary
  channels. 
 \item {\em Seeding, accretion and merging} of black holes is
  implemented following \cite{Springel05b,Booth09}, modified to account for the angular momentum of
  accreted gas as described by \cite{Rosas15}.
 \item Thermal feedback from stars is implemented as described by \cite{DallaVecchia12}; feedback from accreting black holes is also implemented thermally.	
\item {\em Dark matter haloes} are identified using the friends-of-friends 
  algorithm, with baryonic particles (gas, stars, and black holes) assigned 
  to the same halo as the nearest dark matter particle, if any. 
  The mass of the halo is characterised by $M_h\equiv M_{200,c}$, the mass 
  enclosed within a sphere within which the mean density is 200 times the 
  critical density.   
\item {\em Galaxies} are identified using the {\sc subfind} algorithm 
  \citep{Springel01, Dolag09}.  To avoid including \lq intra-cluster\rq\ mass/light to massive
  galaxies, we calculate (and quote) galaxy stellar masses/luminosities
  within 3D spherical apertures of 30 kpc. The aperture size was chosen to broadly approximate 
  a Petrosian aperture, see \cite{Schaye15} for details.
  We classify the galaxy that contains the particle with the lowest potential 
  as the \lq central galaxy\rq, any other galaxies in the same halo is a
  \lq satellite\rq. 
  
\item {\em Broad-band absolute magnitudes} of galaxies are computed 
in the rest-frame, as described by \cite{Trayford15}: stellar emission
is represented by the \cite{bc03} population synthesis models, with dust
accounted for using the two-component screen model of 
\cite{Charlot2000}. Dust-screen optical depths depend on the mass of enriched,
star-forming gas in galaxies and include an additional scatter to represent 
orientation effects.  This constitutes the fiducial model of \cite{Trayford15}
(referred to as GD+O in that paper).
\end{itemize}

The analysis of the \eagle\ simulations is greatly simplified by
using the {\sc SQL} database described by \cite{McAlpine16} 
that contains all the properties of \eagle\ galaxies used here. 
In particular, we extract position, velocity, stellar mass, star formation rate,
and broad-band luminosities in a 30~kpc aperture for all galaxies from the database,
and then convert them to $h$ dependent units (such as $h^{-1}~$Mpc for lengths and $h^{-2}$~M$_\odot$
for stellar masses). 

Galaxies in \eagle\ show similar colour bi-modality as those in \gama\ 
\citep[][and Trayford {\em in prep.}]{Trayford15}: a blue cloud of star forming galaxies and a red sequence of mostly passive galaxies. In the simulation, the appearance of passive red galaxies is due to the suppression of star formation in satellites and by feedback from supermassive black hole, as demonstrated by \cite{Trayford16}. The colour cut used by \cite{Farrow15} to separate red from blue galaxies in \gama\ is (their Eq.~4)
\begin{equation}
	(g-r)_0 = 0.618 - 0.03\,(M_{r,h} + 18.6)\,,\,\,\,\,\hbox{(\gama\ red-blue cut)}\,,
	\label{eq:col_split_gama}
\end{equation}
shown as a dashed black line on the rest-frame  colour-magnitude plot of Fig.~\ref{fig:col-mag}. Here we separate \eagle\ galaxies in a red and blue population using

\begin{equation} (g - r)_{0} = 0.7 - 0.028 \, (M_{r,h} + 18.6)\,\,\,\,\hbox{(\eagle\ red-blue cut)}\,,
\label{eq:col_split_eagle}
\end{equation}
shown as the black line in Fig.~\ref{fig:col-mag}. 
Therefore, the aim of the colour cut is to separate the passive from the active population of galaxies. 
It is essential to ensure that comparable cuts are made in the different data sets.
The slope of observed and simulated colour cuts are virtually identical, and they are offset in colour by less than 0.1 magnitude at $M_{\rm r,h}=-18.6$. This offset is comparable to the offset of $\sim0.15$ (but in the opposite direction to the \eagle\ offset) for the semi-analytical model considered by \citet{Farrow15} (see their Fig.~2).

\subsection{Dark Matter only simulations}
\label{sec:sim_dm}

We use two dark matter only simulations to study the impact of the limited simulation volume  of the \eagle\ simulation on clustering, one with the same volume and initial conditions as \eagle\ 
\citep[referred to as \simdmo\ below, this is the simulation L00100N1504-DMO described by][]{Schaye15}
and one with a much larger simulation volume (referred to as \pmil\ below). Combined they allow us to asses
     to what extent missing large-scale power and sample variance affect inferences on clustering from the relatively small \eagle\ volume.
     
Simulation \simdmo\ has the same volume, gravitational softening, cosmology and initial conditions as \eagle. The masses of the dark matter particles are increased by a factor of $(\Omega_{\rm b}+\Omega_{\rm dark})/\Omega_{\rm dark}$ compared to \eagle, to account for not including the baryonic mass. We use this simulation to study clustering of haloes compared to other models - without results being affected by galaxy formation. The impact of baryonic effects on the density profiles of haloes in \eagle\ was investigated by \citet{Schaller15}.

The \pmil\ simulation (see Table~\ref{tab:sims}; Baugh~et~al.~{\em in prep.} and McCullagh~et~al.~{\em in prep.} ) 
uses identical cosmological parameters as \eagle\ but has a much larger 
volume (800$^3$Mpc$^3$ compared to the 100$^3$Mpc$^3$ for the \eagle\ 
simulation used here). With \pmil\ we can quantify the effects of missing 
large-scale power and poor sampling of long wavelengths on clustering 
statistics by comparing the dark matter halo clustering in \simdmo\ 
with \pmil. For details about \pmil\ see Baugh~et~al.~{\em in prep} and McCullagh~et~al.~{\em in prep}.

In both \simdmo\ and \pmil, dark matter haloes were identified using the friends-of-friends (FoF) algorithm with the standard value of $b=0.2$ for the linking length in units of the mean particle separation. The mass of the halo is represented by $M_{200,c}$, the mass enclosed within a sphere with a density 200 times the critical density.
\subsection{The \gama\ survey}

To put the \eagle\ simulation results into context, we will compare
them primarily to results from the \gama\ survey data, and in particular to clustering
measurements made by \cite{Farrow15}.
The Galaxy And Mass Assembly (\gama) survey \citep{Driver2011,Liske2015} is a spectroscopic and
multi-wavelength survey of galaxies carried out on the Anglo-Australian
telescope. In this work, we make use of the
main r-band limited data from the \gama\ equatorial regions ($\sim$180
sq.deg.), which consists of a highly complete ($>98$~per cent) spectroscopic
catalogue of galaxies selected from the SDSS DR7 \citep{Abazajian09} to 
$r_{\rm petro}<19.8$. Further details of the \gama\ survey input
catalogue, tiling algorithm, redshifting and survey progress are
described by \cite{Baldry2010}, \cite{Robotham2010}, \cite{Baldry2014} and
\cite{Liske2015}, respectively.

For the clustering comparisons presented in \S~\ref{sec:results}, we
are primarily interested in the following \gama\ galaxy properties:
r-band absolute magnitude, stellar mass and rest-frame $(g-r)_0$ colour.
We describe in turn how each of those
properties have been estimated in the clustering measurements of
\cite{Farrow15}.
\begin{itemize}
\item {\em r-band absolute magnitude} ($M_{r,h}$): 
we apply evolution corrections and k-corrections to $z_{\rm ref}=0$ 
  Petrosian r-band absolute magnitudes, 
  where $M_{r,h}\equiv M_r-5\,log_{10}(h)$. For further details, 
  see \S~2.1.4 of \cite{Farrow15}.
\item {\em $(g-r)_0$ colour}: the rest-frame colours are derived from
  SDSS model magnitudes, with colour and redshift dependent
  k-corrections as per \cite{Loveday2012} and
  \cite{McNaught-Roberts2014}. For further details, see \S~2.1.4 of
  \cite{Farrow15}. 
\item {\em stellar mass} ($M_\star$ in units of $h^{-2}$M$_{\odot}$):
  the clustering measurements in \cite{Farrow15} use the relation between
  rest-frame $(g-i)_0$ colour and stellar mass as derived by
  \cite{Taylor2011}, using the \cite{bc03} synthetic stellar population
  models with a \cite{Calzetti00} dust attenuation law to correct for dust in the Milky Way. 
  For further details, see \S~2.1.5 of \cite{Farrow15} and  \cite{Taylor2011}.
\end{itemize}

For completeness, we refer the reader to \S~3 of \cite{Farrow15}  for the modelling of the \gama\ selection function. Accurate modelling of the selection function of a redshift survey is key for precise clustering measurements. The modelling approach described in \cite{Farrow15}, which is based on the method of \cite{Cole2011}, 	enables a uniform modelling for all galaxy samples split by stellar mass, luminosity and colour.

\section{Galaxy clustering}
\label{sec:methods}
In this section we present the analysis methods used, starting with
the estimators we use for calculating the two-point correlation function and its
associate errors. We then briefly discuss how we compute the effective bias.
 
\subsection{The two-point correlation function}

The spherically averaged two-point correlation function, $\xi(r)$,
defined as \citep[e.g.][]{Peebles1980} 
\begin{equation}
\xi(r) = {1\over\langle n\rangle}\,{{\rm d}P\over {\rm d}V} -1\,,
\label{eq:xi1}
\end{equation}
provides a statistical description of a sample's spatial
distribution. Here, ${\rm d}P/{\rm d}V$ is the probability of finding
a galaxy in the volume ${\rm d}V$ at a given (co-moving) distance 
$r$ from another
galaxy, and $\langle n\rangle$ is the mean (co-moving) number density of
galaxies. In practice for a volume with periodic boundary conditions,
the correlation function can be estimated by counting the number of
pairs of galaxies, $N_s(r)$, in a shell of volume $V_s(r)$ at distance
$r$ from each other using \citep[e.g.][]{Rivolo1986} 
\begin{equation}\label{CF_eq}
\xi(r) = \frac{1}{\langle n\rangle^2 V} \frac{N_{s}(r)}{V_{s}(r)} - 1\,,
\end{equation}
where $V$ is the total volume of the periodic simulation. 

However, when the volume has boundaries (as is the case for any observed 
survey, or when we want to restrict the analysis to a small region within a larger periodic simulation volume), these equations cannot be used. In such cases, $\xi$ can be computed by 
comparing the distribution of galaxy pairs to the clustering of a set of 
points uniformly distributed within the survey volume, using e.g.\ the 
\cite{Landy1993} estimator:
\begin{equation}
  \xi(r) = \frac{DD - 2 DR + RR}{RR}\,,
  \label{eq:xi_LS}
\end{equation}
where $DD$ is the suitably normalised number of galaxy pairs at a 
distance $r$ from each other, $RR$ the corresponding normalised number
of pairs from the random distribution, and $DR$ the suitably
normalised numbers of galaxies and random pairs separated by distance $r$.

The co-moving distance between galaxies cannot be measured directly 
from a redshift survey due to galaxy peculiar velocities and large-scale 
redshift space distortions. However, by splitting the information
into projected separation, $r_p$, and distance parallel to the line of 
sight, $\pi$, one can estimate the 2D correlation function, 
$\xi(r_{\rm p},\pi)$, which in turn is used to estimate the projected 
correlation function, 
\begin{equation}
  w_p (r_{\rm p}) = 2 \int_{0}^{\pi_{\rm max}} \xi(r_{\rm p}, \pi) d\pi\,,
  \label{eq:wp_rp}
\end{equation}
with $\pi_{\rm max}$ set to a value adequate for the sample considered (here
$\pi_{\rm max}$ is fixed to $\sim34$~Mpc/$h$, which represents $\sim L/2$ of \eagle\ simulation. See Table~\ref{tab:sims}).
We select $\pi_{\rm max}$ 
to be sufficiently large to account for most redshift space distortions.
In addition the $\pi_{max}$ value chosen is in line with what is commonly used in observational clustering measurements.
To a very good approximation, $w_p (r_{\rm p})$ is independent of redshift 
space effects, making this statistic ideal for model comparisons. 
Furthermore, we tested the systematic differences between $\xi(r)$ and $w_p (r_{\rm p})$
and their dependence on $\pi_{max}$, finding that the systematic difference is significantly smaller than the
statistical errors.
We compute $w_p(r_{\rm p})$ along three orthogonal directions in the simulations, to improve 
the signal-to-noise of the clustering measurements.

To reduce the dynamical range when plotting $w_p$, we will often divide by the projected correlation
function of the reference power law, $\xi(r) = (r_0/r)^{\gamma}$ with $r_{0}=5.33 h^{-1}{\rm Mpc}$ and $\gamma=1.8$, 
 from \citet{Zehavi2011} for the galaxy sample with $-21.0 < M_{r,h} < -20.0$, and where the constants are from the fit
function that corresponds to this power-law is
\begin{eqnarray}
  w_{p}^{\rm ref}(r_{\rm p}) & = & r_{\rm p} \big(\frac{r_{\rm 0}}{r_{p}}\big)^{\gamma} \frac{\Gamma(1/2) \Gamma((\gamma -1)/2)}{\Gamma(\gamma/2)}\,,
  \label{eq:wp_ref}
\end{eqnarray}
where $\Gamma$ denotes the Gamma function. 

\subsection{Error estimates on clustering statistics}
We compute and quote jackknife errors on the simulated two-point correlation function in \eagle\ to mimic observational errors. However, sample variance is likely to dominate the error budget. We estimate sample variance by sub-sampling \eagle\ sized-volumes in the \pmil\ simulation, which also allows us to examine any effects due to missing large-scale power. Unfortunately
we can only estimate these errors for the clustering of haloes - not galaxies - since the
\pmil\ simulation is dark matter only. 

We apply the jackknife technique by partitioning the \eagle\ (or \simdmo) simulation volume in $N_{\rm sub}$ tiles of equal volume, with $N_{sub}=8$. We then compute the two-point correlation function $\xi^{JK}_{k}$ by omitting the $k$-th tile, and compute the variance 
\begin{equation}
  \sigma^{2}(r)  = \frac{(N_{\rm sub}-1)}{N_{\rm sub}} \sum_{k=1}^{N_{\rm sub}}  ( \xi^{JK}_{k}(r) - \xi^{\rm tot}(r) )^2\,,
\end{equation}
\noindent where $\xi^{\rm tot}(r)$ is the correlation function of the total volume. Such jackknife error estimates have been used extensively to estimate errors in galaxy clustering 
\citep[e.g.,][]{Zehavi2002,Zehavi2011,Favole2016}, and \cite{Zehavi2002} shows that such errors
accurately reflect uncertainties in the clustering on the scales investigated here. However, it has of course its limitations, as pointed out by e.g. \citet{Norberg2009}. For example the technique may underestimate errors when a few systems dominate the signal. We will see below this is in fact the case in \eagle, where the clustering of low-mass red galaxies on small scales is dominated by satellites in a few massive clusters, as we illustrate in Fig.~\ref{fig:red-blue-GAMA} below.

We estimate errors on the clustering of haloes due to sample variance and missing large-scale power on \eagle\ volumes using the \pmil\ simulation as follows. We partition the \pmil\ simulation in $N_{\rm sub}=512$ tiles of volume equal to that of \eagle. We then calculate the correlation function of haloes in the $i$-th tile, $\xi_i(r)$, using Eq.~(\ref{eq:xi_LS}), as well as the correlation function of the total volume, $\overline\xi(r)$. The variance is then calculated as
\begin{equation}
\label{eq_PM}
 \sigma^{2}(r) = \frac{1}{{N_{\rm sub}-1}} \sum_{i=0}^{N_{\rm sub}} (\xi_{i}(r) - \overline{\xi(r)})^{2}\,.
\end{equation}
\noindent We use the number density of haloes of each non-overlapping tile to compute $\xi_i(r)$.

\subsection{Effective bias}
The bias, $b$, of a tracer population is the ratio of the correlation 
function of that tracer over that of the mass \cite[e.g.][]{Davis85},
with the effective bias, $b_{\rm eff}(X)$, given by \cite[e.g.][]{Porciani2004}

\begin{equation}
  b_{\rm eff}(X) =  \frac{\int b(M_h) \, N_{gal}(M_h,X) \, n(M_h) \, dM_h}{\int  N_{gal}(M_h,X) \, n(M_h) \, dM_h},
  \label{eq:eff_bias}
\end{equation}
where $n(M_{h})$ is the halo mass function (the number density of haloes of mass $M_h$), 
$b(M_{h})$ the linear bias factor of haloes of mass $M_h$, and $N_{\rm gal}(M_h,X)$ the mean 
number of galaxies of property X in haloes of mass $M_h$ (the mean halo occupation). The property $X$ could
select galaxies in a given stellar mass, colour or luminosity range, for example. We 
approximate the integral as a sum over all haloes in the simulation, obtaining
\begin{equation}
  b_{\rm eff}(X) = \frac{\sum_{i=0}^{N_{\rm haloes}} b(M_h^i) N_{\rm gal}(M_h^i,X)}{\sum_{i=0}^{N_{\rm haloes}} N_{\rm gal}(M_h^i,X)}\,,
 \label{eq:eff_bias2}
  \end{equation}
where $N_{\rm gal}(M_h^i,X)$ is the number of galaxies with property X 
in a halo of mass $M_h^i$. In practice we estimate the effective bias of 
samples split by stellar mass and hence evaluate this sum for all galaxies 
in narrow stellar mass bins.

To estimate the linear halo bias $b(M_h)$, we follow \citet{Mo1996},
\begin{equation}
  b(M_h) = 1 + (\nu(M_h)^2 - 1)/\delta_{c} \,,	
\end{equation}
where $\delta_{c} = 1.686$ is the spherical collapse density threshold, 
and $\nu(M_h) = {\delta_{c}}/{\sigma(M_h)}$ the dimensionless amplitude of 
fluctuations that produce haloes of mass $M_h$ (at a given redshift $z$). 
The (linear) matter variance, $\sigma(M_h)$ at a given $z$, can be computed 
numerically for a given linear power spectrum \citep[see Eq.~(9) in][]{murray13} using 
the web-portal HMFcalc\footnote{http://hmf.icrar.org} (and adopting the spectral 
index $n_s = 0.9611$ used in \eagle). Finally, we adopt the 
fit provided by \cite{Jenkins2001} for the halo mass function, $n(M_h)$, which 
provides a good description of the halo mass function of all simulations used here.

\section{Results}
\label{sec:results}

We begin by considering the clustering of 
dark matter haloes in \eagle, followed by a quick look at 
the real and redshift space clustering of \eagle\ galaxies 
with well resolved stellar mass. In \S\ref{sec:gama_vs_eagle}
we present the main results of this study, namely the clustering 
of galaxies in \eagle\ compared to that of \gama, when split by 
stellar mass, luminosity, colour or star-formation rate.

\begin{figure}
  \centering
  \includegraphics[width=\columnwidth]{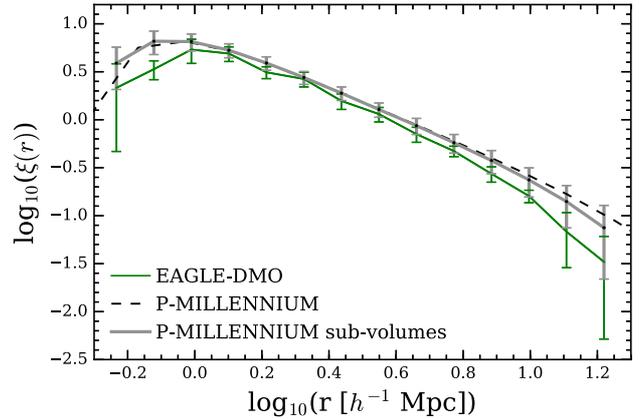}
  \caption{The $z=0$ real space two-point correlation function 
    of haloes with mass $M_h > 10^{12} {\rm M}_{\odot} h^{-1}$ from \pmil\ 
    (\textit{black dashed line}), the mean correlation function, $\overline\xi(r)$, of the 512 
    non-overlapping tiles of that simulation (\textit{grey line}), and the 1$\sigma$ scatter around this mean 
    (computed with Eq.~\ref{eq_PM}, \textit{grey error bars}). 
    The \textit{green curve} is the correlation function for simulation \simdmo\ with
    jackknife error bars. Finite-volume effects cause the green curve to fall 
    increasingly below the grey line, with jackknife errors yielding 
    nevertheless a realistic error estimate.
    }
 \label{fig:xi_halo}
\end{figure}

\subsection{Halo clustering in EAGLE}
\label{sec:clus_dm}
The real-space clustering of haloes with $M_h > 10^{12} \,h^{-1}{\rm M}_{\odot}$ 
in the \pmil\ dark matter simulation is plotted in Fig.~\ref{fig:xi_halo}. 
We sub-divide the volume of this simulation into 512 non-overlapping tiles, 
each with the same volume as \simdmo, and compute the correlation function $\xi_i(r)$ for each 
of the tiles, using the mean number density of haloes in each tile in Eq.(\ref{eq:xi_LS}).
We plot the mean correlation function averaged over all tiles, $\overline\xi(r)$, as the grey line, 
with the scatter around the mean shown as $1\sigma$ error bars. The mean correlation function
$\overline\xi(r)$ follows the correlation function of the full volume (black dashed line) very closely,
falling well within the scatter between volumes, as expected.

The correlation function of \simdmo\ (green line) falls below $\overline\xi(r)$
on scales smaller than $1~h^{-1}$~Mpc, remains well within sample variance up 
to scales $5-6~h^{-1}$~Mpc ($\log r/(h^{-1}{\rm Mpc})=0.7-0.78$), then falls 
increasingly below $\overline\xi(r)$ above this scale. As both simulations have 
identical power-spectra and cosmological parameters, the deviations are due 
to sample variance and to the integral constraint on $\xi$. We note that numerical resolution is not likely 
to play a role in the apparent differences in clustering, as these haloes are resolved by $\approx 10^4$ particles 
or more.

We compute jackknife errors for \simdmo, as described above, and plot them in green.
Of course these do not quantify finite-volume effects nor sample variance, but 
nevertheless the green and black curves are within the \simdmo\ jackknife errors. 
This motivates us to use such errors when calculating errors on the correlation 
function of galaxies, rather than haloes, below, since we do not have access to larger
hydrodynamical simulations to estimate finite-volume effects.
Given the level of convergence between \simdmo\ and \pmil, we will plot 
correlation functions up to scales of 10$~h^{-1}$~Mpc (14~per cent of the full extent), 
with Fig.~\ref{fig:xi_halo} quantifying the limitations on halo clustering.

\subsection{Galaxy clustering in \eagle}
\label{sec:clus_gal}

\begin{figure}
   \centering
   \includegraphics[width=\columnwidth]{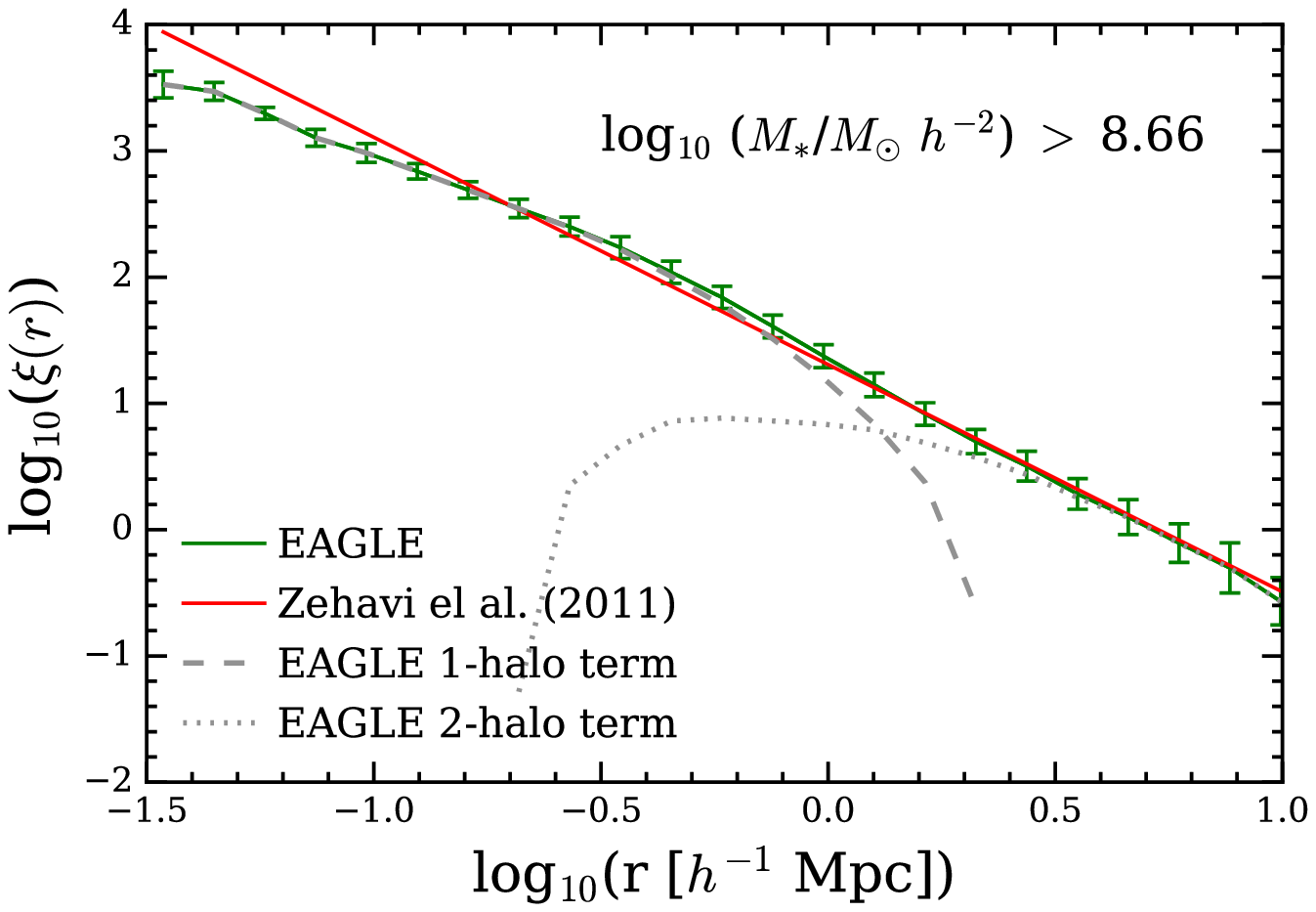}
   \includegraphics[width=\columnwidth]{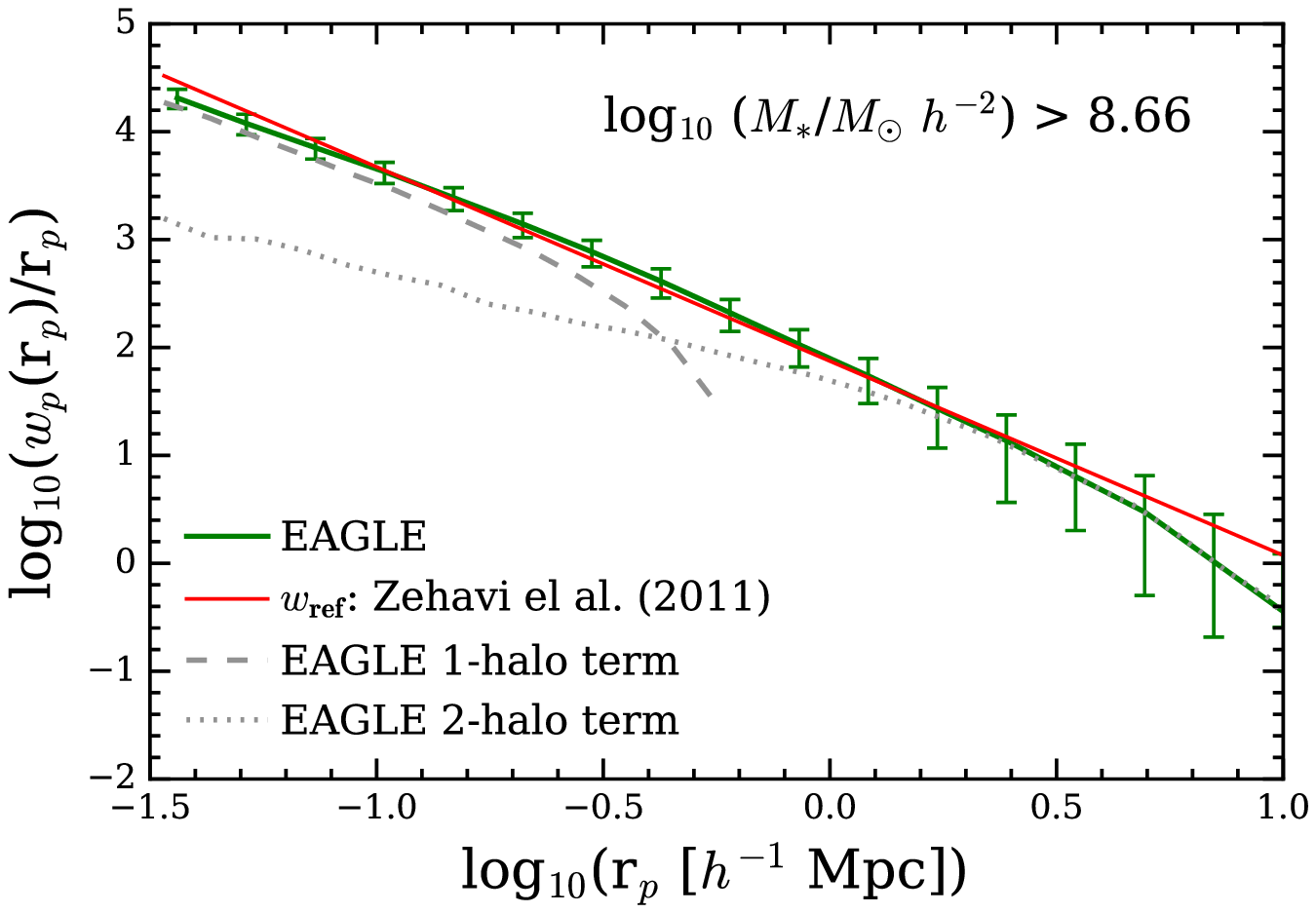}
   \caption{\textit{Top panel:} the $z=0.1$ two-point correlation function for \eagle\ 
   	galaxies with $M_\star>10^{8.66} h^{-2} {\rm M}_{\odot}$ (\textit{green curve with jackknife
   	errors bars}), decomposed in the one- and two-halo terms (\textit{dashed} and \textit{dotted grey
   	lines}, respectively). The \textit{red line} is the reference power law model	 for galaxies with
   	$-21.0 < M_{r,h} < -20.0$ from the fit by \citealt{Zehavi2011}. 	\textit{Bottom panel:} 
   	corresponding projected correlation function, $w_p(r_p)$ from Eq.~(\ref{eq:wp_rp}). 
   	The \textit{red line}
   	is the projected correlation function from the fit by \citealt{Zehavi2011},
   	$w_p^{\rm ref}$ from Eq.~(\ref{eq:wp_ref}). 
   	The galaxy selection
   	is different in detail for \eagle\ and the observations: the red lines are shown to guide the eye.}
   \label{fig:xi_mstar}
 \end{figure}

\begin{figure}
  \centering
  \includegraphics[width=0.8\columnwidth]{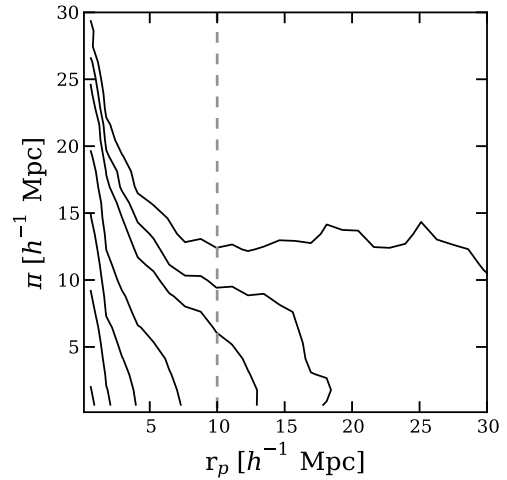}
  \caption{The $z=0.1$ two-dimensional redshift space correlation function $\xi(r_{\rm p},\pi)$, 
    as function of projected separation $r_{\rm p}$ and line-of-sight separation 
    $\pi$, for \eagle\ galaxies with stellar masses greater than 
    $10^{8.66} h^{-2} {\rm M}_{\odot}$. Contours levels correspond to 
    $\xi(r_p,\pi) = 5, 2, 1, 0.5, 0.2, 0.1, 0.01$.
    The \textit{vertical dashed line} corresponds to projected separations of $r_p = 10 h^{-1}{\rm Mpc}$ beyond which the clustering of haloes is increasingly suppressed due the limited 
    extent of the \simref\ simulation (see Fig.~\ref{fig:xi_halo}).}
  \label{fig:xi_2d}
\end{figure}

The real-space correlation function, $\xi(r)$, of galaxies with stellar mass 
$M_\star>10^{8.83} h^{-2} {\rm M}_{\odot}$ from \simref\ is plotted in 
Fig.~\ref{fig:xi_mstar} (top panel). Distinguishing between central galaxies 
(typically but not necessarily the most massive galaxy in a given halo)
and satellites galaxies, we compute the one- and two-halo contributions 
separately (grey dashed and grey dotted lines, respectively). The contribution to $\xi$ from these is 
equal at a separation of approximately $r=1.3~h^{-1}$~Mpc. 
It is important to note that, while the two-point correlation function from the dark
matter haloes of \simdmo\ is underestimated at distances below $r\sim 1\,h^{-1}{\rm Mpc}$,
this does not imply that the same is true for the galaxy correlation function, since there are
of course generally many galaxies per halo. 

The real-space correlation function quantifies the physical clustering of galaxies, independently of any peculiar velocities. 
However, observations can only measure clustering in redshift space, and peculiar velocities then distort the signal. To ameliorate the effects of such redshift space distortion, it is convenient to integrate $\xi(r_p,\pi)$ over a narrow range in $\pi$, and compute the projected correlation function $w_{\rm p}(r_{\rm p})$, see Eq.(\ref{eq:wp_rp}). This is plotted for the same \eagle\ galaxies (those with $M_\star>10^{8.83} h^{-2} {\rm M}_{\odot}$) in the bottom panel of Fig.~\ref{fig:xi_mstar}, again plotting one- and two-halo terms as well.

The reference model $w_p^{\rm ref}(r_{\rm p})$ of Eq.~(\ref{eq:wp_ref}) provides a relatively good fit to \eagle's projected correlation function. We note, however, that the galaxy selections differ between \eagle\ and the {\sc sdss} galaxies fit by \citet{Zehavi2011} that gives rise to $w_p^{\rm ref}$ -- the two therefore did not have to agree: we show the comparison to guide the eye, and because we use $w_p^{\rm ref}$ as a normalisation below.

The contribution of one- and two-halo terms to $w_p$ are equal at a projected separation of 
$r_p \sim 0.4 h^{-1}{\rm Mpc}$. Comparing top and bottom panel from Fig.~\ref{fig:xi_mstar},
it is clear that the two-halo term contributes more to the the projected correlation function on scales comparable to 
the virial radius of haloes than it does to the two-point correlation function: at a scale of $\approx 0.1h^{-1}{\rm Mpc}$, the two-halo term contributes nearly 10~per cent to $w_p$.

The two-dimensional redshift space correlation function of this stellar
mass limited sample of \eagle\ galaxies is plotted in Fig.~\ref{fig:xi_2d} 
in terms of the projected separation, $r_p$, and line of sight separation, $\pi$. 
It exhibits the familiar elongation in the $\pi$ direction at small $r_p$ 
resulting from virial motion of galaxies in haloes (the \lq fingers-of-god\rq\ 
effect), and the flattening in the $\pi$ direction at large $r_p$, due to 
coherent streaming motions of galaxies into haloes and out of voids 
(the \lq Kaiser\rq\ effect, \citet{Kaiser87}). We do not compare this correlation function
directly to \gama, mainly because of the complexity of making sure the selection of galaxies
in the $\pi$ direction is the same in simulation and data.

\subsection{Galaxy clustering in \eagle\ compared to \gama}
\label{sec:gama_vs_eagle}

We use the volume-limited samples of \gama\ galaxies presented by \citet{Farrow15}, which we can split by stellar mass, luminosity or colour. 
For samples split in bins of stellar mass or luminosity only, we refer the
reader to Table~2 of \citet{Farrow15}, while we present in Table~\ref{tab:gama} the 
properties of the additional \gama\ samples used in here, for which we have
computed clustering statistics following the methods outlined by \citet{Farrow15}\footnote{\citet{Farrow15}
uses a flat $\Omega_m=0.25$ cosmology to infer distances for their clustering measurements. 
On the scales considered, this difference in cosmology is totally negligible.}.

Throughout this section, \eagle\ galaxies are selected from the $z=0.1$ snapshot (a redshift close to the median redshift of the \gama\ samples). Some statistics of the \eagle\ samples used are provided in Table~\ref{tab:eagle}. By construction and as explained in Section~\ref{sec:sim}, the galaxy formation model in \eagle\ yields a stellar mass function that is in relatively good agreement with that inferred from \gama\ in the mass range we analyse here. The agreement is not as good as in statistical methods that populate dark matter halos with galaxies, such as {\sc sham} or {\sc hod} for example, and which yield the correct number densities by construction. Interestingly however, \eagle\ predicts a scatter in stellar mass at a given halo mass that depends on halo concentration, although the dependence is not strong enough to explain the full variance \citep{Matthee17}. Such non-linear dependencies are not taken into account in these statistical methods. 

Hence, because the mean number density of galaxies in each bin of stellar mass, agree reasonably well between data and simulation, we do not compare clustering at given number density, as commonly done, but directly compare samples selected by stellar mass - or indeed luminosity.

\begin{figure*}
  \centering
  \includegraphics[width=\textwidth]{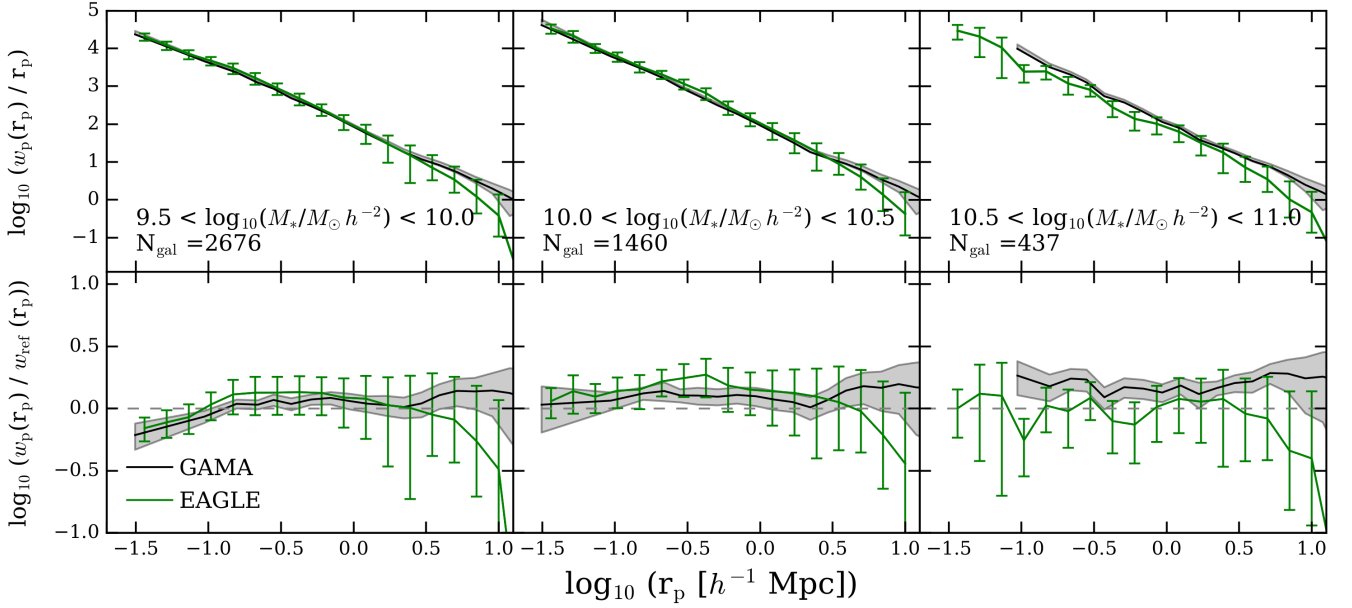}
  \caption{Clustering as a function of stellar mass; the mass bin is indicated in each column.
  \textit{Top panels:} Projected correlation function $w_p(r_p)$ from Eq.~(\ref{eq:wp_rp}). \textit{Green curve} is the \eagle\ result at $z=0.1$, with jackknife error bars; $N_{\rm gal}$ is the number of \eagle\ galaxies in each mass bin. \textit{Black curve} is \gama\ result, with the \textit{grey shading} including the 1-$\sigma$ error range. \textit{Bottom panels:} Same as top panels, but the correlation functions are divided by the reference function $w^{\rm ref}_p(r_p)$ from Eq.~(\ref{eq:wp_ref}). The \textit{dashed line} indicates where the ratio is unity.}
\label{fig:Ms-bin-GAMA}
\end{figure*}

\begin{figure}
  \centering
  \includegraphics[width=0.4\textwidth]{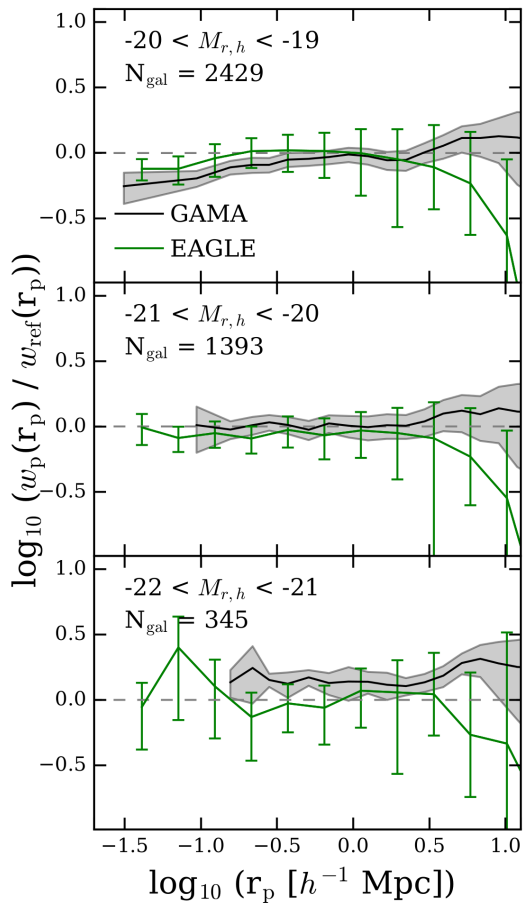}
  \caption{Same as the bottom panels of Fig.~\ref{fig:Ms-bin-GAMA}, but for clustering as a function of $r$-band luminosity. Ranges of $r$-band absolute magnitudes are shown from faint to bright (top to bottom panels), with the number of \eagle\ galaxies in this luminosity range labelled in each panel. The jackknife error bars for \eagle\ galaxies are included.}
 \label{fig:Mr-bin-GAMA}
\end{figure}

\begin{table*} 
  \begin{center}
    \caption{Statistics of \gama\ galaxy samples (mostly volume limited), that are not already described in Table~2 of \citet{Farrow15}, following a similar table structure. The stellar mass restricted
      samples are split into red and blue galaxies, presented in turn. Columns from left to right are: 
      stellar mass range, minimum and maximum sample redshift, 
      total number of galaxies, the galaxy number density, median sample redshift, median $r$-band absolute 
      magnitude, median stellar mass, median $(g-r)_0$ rest-frame colour, and 
      the fraction of truly volume limited galaxies (see text for further details).
     }
    \label{tab:gama}
    \begin{tabular}{lrrrcrcrrr}
      \hline
      stellar mass range & $z_{\rm min}$ & $z_{\rm max}$ & N$_{\rm gals}$ & $\bar{n}$ & $z_{\rm med}$ & M$^{\rm med}_{r,h}$ & $\log_{10}$ M$_*^{\rm med}$ & $(g - r)^{med}_{0}$ & $f_{vlim}$ \\ 
            & & & & $({\rm Mpc}/h)^{-3}$& & & $(h^{-2}\,{\rm M}_{\odot})$ & & \\
      \hline 
      Red   &                &               &                &               &               &                                          &                   &                \\
      \hline
      $9.5<\log_{10} {\rm M}_*/h^{-2}\,{\rm M}_{\odot}<10.0$  & 0.02 & 0.14 & 5407 &  4.45\,10$^{-3}$ & 0.11 & -19.16 (0.38) & 9.77 (0.14) & 0.72 (0.07) & 0.96 \\
      $10.0<\log_{10} {\rm M}_*/h^{-2}\,{\rm M}_{\odot}<10.5$ & 0.02 & 0.14 & 5527 &  4.55\,10$^{-3}$ & 0.11 & -20.16 (0.40) & 10.23 (0.14) & 0.75 (0.10) & 1.00 \\
      $10.5<\log_{10} {\rm M}_*/h^{-2}\,{\rm M}_{\odot}<11.0$  & 0.02 & 0.14 & 1945 & 1.60\,10$^{-3}$ & 0.12 & -21.11 (0.37) & 10.65 (0.12) & 0.77 (0.09) & 1.00 \\
      \hline
      Blue   &                &               &                &               &               &                                          &                   &                \\
      \hline
      $9.5<\log_{10} {\rm M}_*/h^{-2}\,{\rm M}_{\odot}<10.0$  & 0.02 & 0.14 & 4663 & 3.84\,10$^{-3}$ & 0.11 & -19.71 (0.39) & 9.71 (0.14) & 0.54 (0.08) & 1.00 \\
      $10.0<\log_{10} {\rm M}_*/h^{-2}\,{\rm M}_{\odot}<10.5$ & 0.02 & 0.14 & 1870 & 1.54\,10$^{-3}$ & 0.12 & -20.61 (0.34) & 10.18 (0.13) & 0.61 (0.06) & 1.00 \\
      $10.5<\log_{10} {\rm M}_*/h^{-2}\,{\rm M}_{\odot}<11.0$ & 0.02 & 0.14 & 248 & 0.20\,10$^{-3}$  & 0.12 & -21.44 (0.34) & 10.61 (0.11) & 0.67 (0.06) & 1.00 \\
    \end{tabular}
  \end{center}
\end{table*}

\begin{table*}
  \centering 
  \caption{Statistics of \eagle\ stellar mass selected samples from the $z=0.1$ snapshot.
  Columns from left to right are: stellar mass range, number of galaxies, galaxy number density (including JN errors),  
    fraction of satellites, fraction of blue galaxies (following Eq.~\ref{eq:col_split_eagle}), the SFR limits used to define \lq low\rq\ and \lq high\rq\ SFR galaxy samples respectively, number density of red and blue galaxies, respectively.
    }
  \label{tab:eagle}
  \begin{tabular}{lrcrrrrcc}
    \hline
    Sample  & N$_{\rm gals}$  & $\bar{n}$                 & $f_{\rm sat}$ & $f_{\rm blue}$  & SFR$_{\rm low}$            & SFR$_{\rm high}$            & $\bar{n}_{\rm red}$    & $\bar{n}_{\rm blue}$\\
        stellar mass range    &  & $({\rm Mpc}/h)^{-3}$                     &               &                 & ${\rm M_{\odot}\,yr^{-1}}$ & ${\rm M_{\odot}\,yr^{-1}}$  & (Mpc/$h$)$^{-3}$   & (Mpc/$h$)$^{-3}$      \\
    \hline
    9.5$<\log_{10} {\rm M}_*/h^{-2}\,{\rm M}_{\odot}<$10.0 & 2676 & (8.6$\pm$0.9)\,10$^{-3}$ &   0.43  &  0.82  & $<$0.28 & 1.02$<$  & 1.5\,10$^{-3}$ &  7.1\,10$^{-3}$ \\ 
    10.0$<\log_{10} {\rm M}_*/h^{-2}\,{\rm M}_{\odot}<$10.5 & 1460 & (4.7$\pm$0.5)\,10$^{-3}$&   0.39 &  0.77  & $<$0.26 & 1.99$<$  & 1.2\,10$^{-3}$ &  3.6\,10$^{-3}$ \\ 
    10.5$<\log_{10} {\rm M}_*/h^{-2}\,{\rm M}_{\odot}<$11.0 & 437 & (1.4$\pm$0.1)\,10$^{-3}$ &   0.22 &  0.81  & $<$0.65 & 3.43$<$  & 2.7\,10$^{-4}$ &  1.1\,10$^{-3}$ \\ 
    \hline
   \end{tabular}
\end{table*}

\subsubsection{Stellar mass dependent clustering}

A comparison of the clustering of \eagle\ galaxies to that in \gama\ as a function of stellar mass is shown in Fig.~{\ref{fig:Ms-bin-GAMA}, using the same mass as used by \citet{Farrow15}. The bottom panels of Fig.~\ref{fig:Ms-bin-GAMA}
highlight the differences between the two measurements, by presenting
the ratio of the projected correlation functions with respect to the
reference power law model adopted 
\citep[following][]{Farrow15}. The projected correlation function of 
\eagle\ galaxies (green lines with jackknife errors) is in remarkably good agreement 
with the \gama\ data (solid black lines, with 1-$\sigma$ uncertainty range 
shown as a grey shaded area): the deviations are typically within the 
measured uncertainty range. 
 
It is well known (and clear from the figure) that the clustering strength of galaxies increases with stellar mass. There is little or no evidence for such a trend in the simulation, however the errors are relatively large and the simulation is not inconsistent with such a trend either.
Furthermore, the size of the simulation prevents us to test the stellar mass dependent clustering, 
due to observed trends are only visible when having a large dynamic range of stellar masses.
Therefore, a larger volume would be needed to confirm such trend. The good
agreement in shape of the correlation functions of \gama\ and \eagle\ is encouraging.

The calibration of sub-grid parameters in \eagle\ was based on one-point statistics, as described in Section~2. The clustering of galaxies is therefore a genuine model prediction. The good agreement then implies that \eagle\ galaxies tend to inhabit haloes in a way that mimics accurately the way \gama\ galaxies do.
Finally we note that the decrease of the clustering signal on scales greater than 5~$h^{-1} {\rm Mpc}$ scales in \eagle\ is related to the limited simulation box size - and mimics the corresponding fall in clustering of \eagle\ haloes.

\subsubsection{Luminosity dependent clustering}

A comparison of the clustering of \eagle\ galaxies to that in \gama\ as a function of $r$-band luminosity, is shown in Fig.~\ref{fig:Mr-bin-GAMA}. The agreement is very good, and well within the relatively large jackknife  error estimates. Similar to the case of clustering as function of mass, the amplitude of observed clustering increases with luminosity 
\citep[see e.g.][]{Norberg2001,Zehavi2005}, but again there is little or no evidence for such a trend in \eagle. As in the previous section, we suggest this is mostly due to the finite volume of the simulation: more luminous galaxies are biased to more massive haloes, of which there are relatively few in the \eagle\ volume, and their clustering is underestimated because of lack of large-scale power. The reference power law $w_p^{\rm ref}$ from Eq.~(\ref{eq:wp_ref}), describes the clustering of \eagle\ galaxies in the middle panel of Fig.~\ref{fig:Mr-bin-GAMA} very well. In this panel, \eagle\ galaxies are selected in the same way, $-21.0 < M_{r,h} < -20.0$, as in the sample of \cite{Zehavi2011} to which $w_p^{\rm ref}$ was fit, so they can be
compared directly. The good agreement in clustering, combined with the fact that \eagle\ also fits the galaxy luminosity
function well \citep{Trayford15}, implies that \eagle\ galaxies form in similar haloes, and have similar stellar
populations and star formation histories as those in \gama. This encourages us to look in clustering as function of
galaxy colour in more detail next.
	
\subsubsection{Colour dependent clustering}
\begin{figure*}
  \centering
  \includegraphics[width=0.8\textwidth]{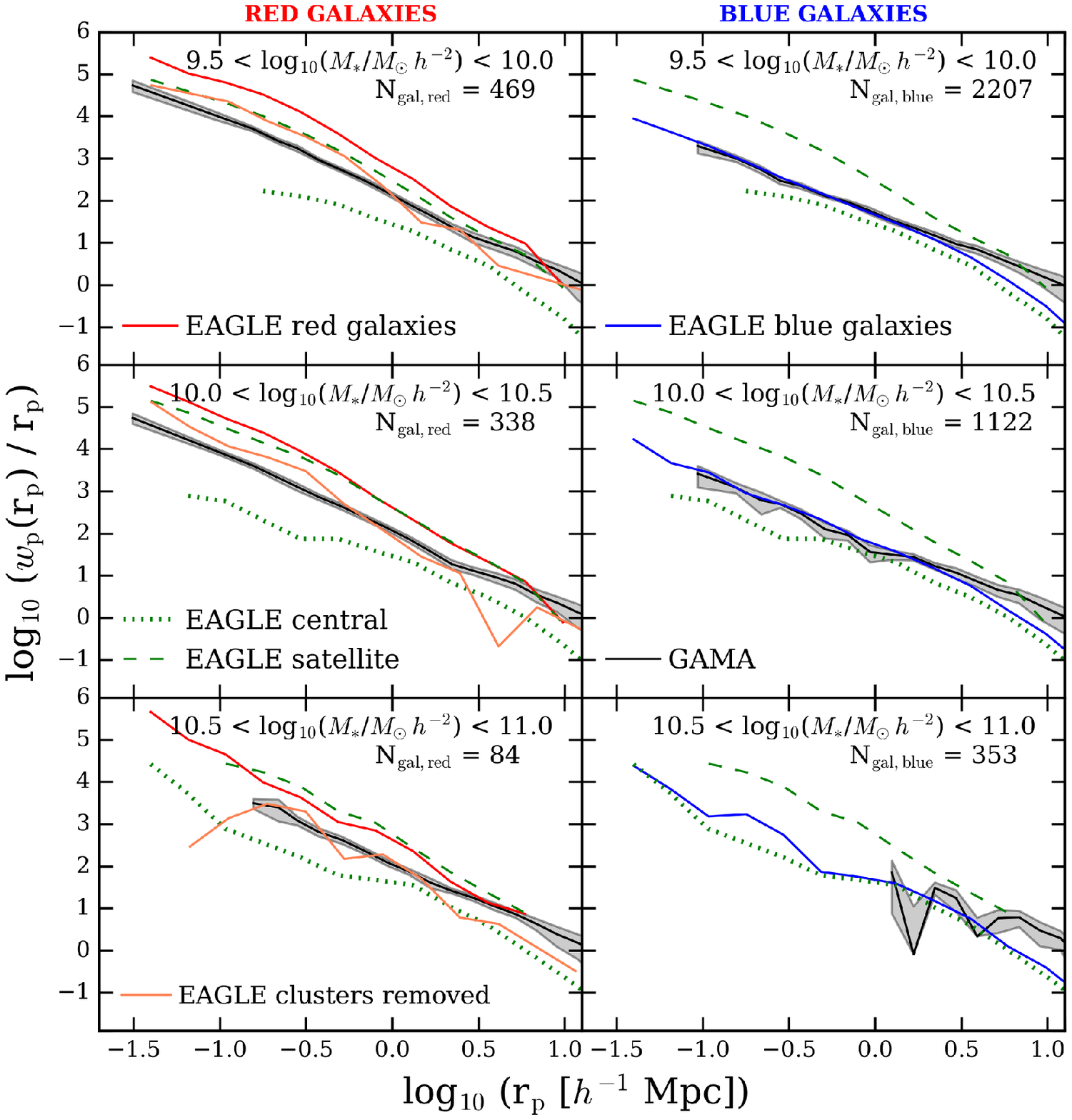}
  \caption{Clustering as function of $(g-r)_0$ colour for red and blue galaxies (left and right column, respectively), in bins of stellar mass (top to bottom rows). Limits of each mass bin are labelled in each panel. The projected correlation function for \eagle\ galaxies in each mass bin is shown by the \textit{red} and \textit{blue} curves for red and blue galaxies, respectively, $N_{\rm gal}$ is the number of \eagle\ galaxies that contributes to the calculation. The corresponding clustering of \gama\ galaxies is shown by the \textit{black line} with \textit{grey shaded} region the 1-$\sigma$ error range. The clustering of central and satellite \eagle\ galaxies split by mass but not by colour, is plotted as \textit{dotted green} and \textit{dashed green lines}, respectively. In the left column, the \textit{orange lines} are $w_p$ of red \eagle\ galaxies, after excluding the three most massive haloes.
      }
  \label{fig:red-blue-GAMA}
\end{figure*} 

\cite{Farrow15} present $r$-band magnitude limited samples of \gama\ galaxies split by rest-frame $(g-r)_0$ colour and in bins of stellar mass. Galaxies are classified as \lq red\rq\ or \lq blue\rq\ using the $r$-band magnitude colour cut of Eq.~(\ref{eq:col_split_gama}). We use this data to compute $w_p$ for these galaxies using the method described by \cite{Farrow15}. Some statistics of these samples are summarised in Table~\ref{tab:gama}. We note that the lowest mass bin in red galaxies, $9.5<\log_{10} {\rm M}_*/h^{-2}\,{\rm M}_{\odot}<10.0$, is only partially volume limited, as $\sim4$ per cent of the galaxies in that sample are only volume limited over a smaller redshift range than the one nominally considered. Given the small galaxy fraction affected by this, we can consider this sample still to be volume limited when computing clustering statistics. The advantages of keeping the exact same volumes for all stellar mass samples split 
by colour (hence they all sample the same underlying large scale structure)
overcomes this minor subtlety, which is primarily driven by uncertainties in the measured colours and adopted k-corrections for a small subset of galaxies. 

We computed colours of \eagle\ galaxies as explained in Section~\ref{sec:sim}. The colour-magnitude diagram exhibits a blue cloud of star forming galaxies, well separated from a \lq red sequence\rq\ of passive galaxies, as shown by \cite{Trayford15} (see also Trayford {\em et.} al. 2017, submitted). The colour of \eagle's red sequence is slightly bluer than in \gama, which may due to differences in metallicity and/or limitations in the adopted population synthesis models, as discussed by \cite{Trayford15}. When comparing to \gama, we want to study whether the clustering of star forming (blue) galaxies
	differs from that of passive (red) galaxies, at a given mass.
	We therefore divide \eagle\ galaxies in bins of stellar mass and
	colour using the same mass bins as used in the analysis of \gama, but
	applying the slightly different colour cut to distinguish red from
	blue using Eq.~(\ref{eq:col_split_eagle}), as compared to the cut of Eq.~(\ref{eq:col_split_gama}) applied to \gama. Some statistics of
	the
	\eagle\ galaxies are summarised in Table~\ref{tab:eagle},
	including the fraction of \eagle\ galaxies that are satellites.

The projected correlation functions for red and blue galaxies, split in bins of stellar mass, is plotted in Fig.~\ref{fig:red-blue-GAMA}. To ease the interpretation of the \eagle\ clustering results, we show the correlation function for all central and satellite galaxies (i.e.\ irrespective of colour) within each stellar mass bin as green dotted and dashed lines respectively. Therefore, the central/satellite clustering results are the same in the left and right panels, but vary with stellar mass (from top to bottom).

In \eagle, red galaxies (left column) cluster more strongly than blue galaxies (right panel) 
of given mass, and similarly satellite galaxies (dashed lines) cluster more strongly than centrals 
(dotted lines), in all stellar mass ranges studied. It is also apparent that the red population follows 
closely the clustering of satellites, in particular for galaxies with stellar masses greater than 
10$^{10} h^{-2} {\rm M}_{\odot}$. In contrast, blue galaxies follow more closely the clustering of 
centrals, again particularly so for the two more massive galaxy stellar bins.

The clustering of blue \eagle\ galaxies tracks that of blue \gama\ galaxies well, in particular 
on scales up to $r_p\sim$4 $h^{-1}$Mpc ($\log_{10}(r_p)=0.6$). However, red \eagle\ galaxies cluster noticeably more strongly than red \gama\ galaxies. As a consequence, the trend that red galaxies cluster more strongly than blue galaxies of given mass, which is clearly present in \gama, is too strong in \eagle. \cite{Trayford15} noticed that red galaxies are overabundant in \simref\ at low mass, and demonstrated that this is at least partly due to lack of numerical resolution \citep[see the Appendix of][]{Trayford15}. At higher stellar masses, \simref\ yields a too large fraction of blue galaxies instead, plausibly a consequence of the dust screen not suppressing blue light from star forming regions sufficiently (Trayford et al. 2017, submitted). We suspect therefore that it is the overly strong suppression of star formation in small galaxies as they become satellites, most likely as a consequence of lack of numerical resolution, that cause \eagle\ to over predict small-scale clustering of red galaxies. Consistent with this interpretation, we find that the strong clustering of red galaxies in \eagle\ is significantly influenced by the presence of a few massive haloes. To demonstrate this, we re-compute $w_p$ for red \eagle\ galaxies after excluding all galaxies in the three most massive haloes. We show the result by the orange line in Fig.~\ref{fig:red-blue-GAMA}. 
The overall amplitude of the clustering signal of red galaxies is dramatically reduced. A larger
simulation box is likely required to provide detailed insight on the clustering of red massive galaxies.

The colour dependences of the clustering of \eagle\ galaxies of given stellar mass 
is clearly partly due to the relative fractions of satellite and central galaxies. 
In Table~\ref{tab:eagle} we show the fraction of satellites and blue galaxies for 
each stellar mass range. We find that the fraction of satellite galaxies decreases at 
higher masses, while the fraction of red and blue satellites is not strongly dependent 
on stellar mass. 
For example, in the stellar mass range $9.5 < \log M_{*}/{\rm M}_{\odot} h^{-2} < 10.0$ 
we find that $\sim 19$\% are red and the remaining $\sim 81$\% blue, while 43\% of the galaxies are satellites.
The table also shows that the vast majority of galaxies are blue, but also the most of galaxies of the complete sample
in this stellar mass bin are centrals.
Furthermore, the good agreement with \gama\ in the clustering of blue galaxies is consistent with the fact that \eagle\ 
galaxies of given mass cluster similarly to \gama\ galaxies, as shown in Fig.~\ref{fig:Ms-bin-GAMA}. 
A relatively modest improvement of the numerical resolution of the simulation could be 
enough to reproduce the clustering of red galaxies equally well \citep[see,][for further details]{Trayford15}.
However the difference seen could equally be due to the missing large-scale 
power and the impact the few rare objects have in the 100~Mpc \eagle\ volume.

\subsubsection{Star formation rate dependent clustering}

\begin{figure}
\centering
\includegraphics[width=0.4\textwidth]{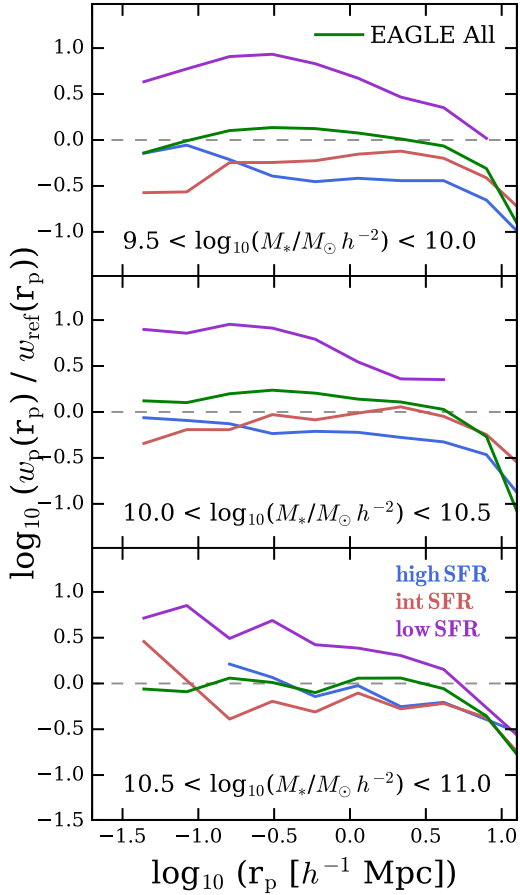}
\caption{Clustering as function of star formation rate. The projected galaxy correlation function of \eagle\ galaxies, divided by the reference power-law of Eq.~(\ref{eq:wp_ref}), is plotted for galaxies
	in bins of stellar mass, as labelled in each panel. In each galaxy stellar mass bin, 
	we show the result for all galaxies (\textit{green line}), the 30~per cent least star forming galaxies 
(\lq low SFR\rq, \textit{purple}), the 30~per cent highest star forming galaxies (\lq high SFR\rq, \textit{blue}), and the remainder (\lq int SFR\rq, \textit{red}). Table~\ref{tab:eagle} lists the corresponding cuts in star formation rate.}
\label{fig:sfr-spl}
\end{figure}

\begin{figure}
 \centering
 \includegraphics[width=\columnwidth]{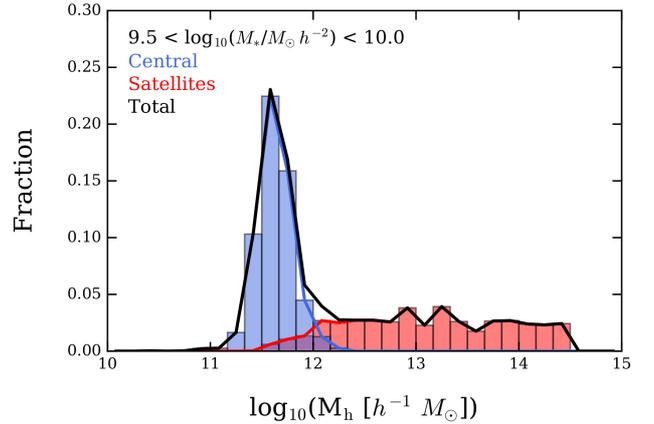}
 \caption{Halo occupation distribution of \eagle\ galaxies with stellar mass in the range
   $9.5 < \log(M_{*}/{\rm M}_{\odot} h^{-2}) < 10.0$. The distribution for all galaxies in that stellar 
   mass range is plotted as the \textit{black} histogram, which is normalised to unit integral. The \textit{red} and \textit{blue} histograms show the fraction of those galaxies that are satellites and centrals, respectively. The coloured histograms integrate separately to the fraction of galaxies that are satellite (red) or central (blue) in this range of $M_\star$.}
 \label{fig:sat-histograms}
\end{figure}

\begin{figure}
  \centering
  \includegraphics[width=0.4\textwidth]{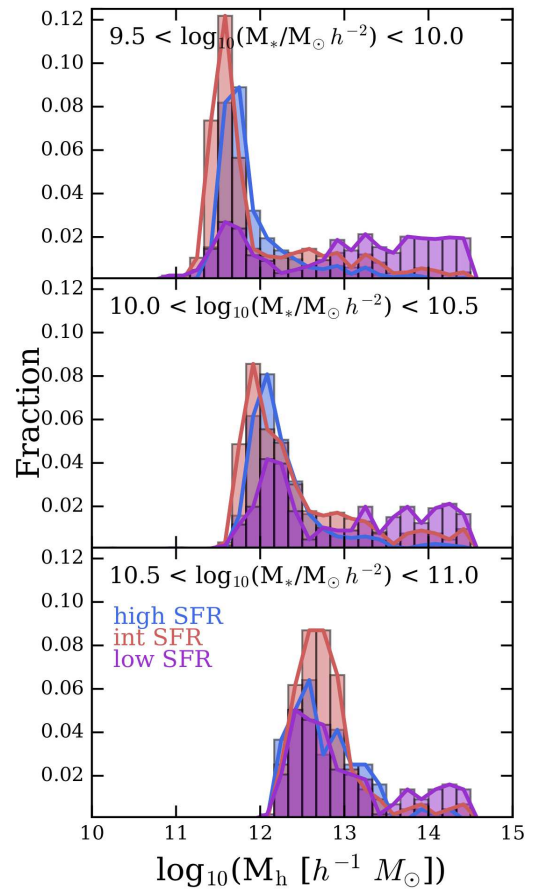}
  \caption{As in Fig.~\ref{fig:sat-histograms}, but for \eagle\ galaxies split by star formation
    rate and for a three ranges in stellar mass as indicated in each panel (with increasing 
    stellar mass from top to bottom). 
    The sample cuts are the same as in Fig.~\ref{fig:sfr-spl} (and listed in Table~\ref{tab:eagle}).
  }
  \label{fig:sfr-histograms}
\end{figure} 

\begin{figure}
  \centering
  \includegraphics[width=\columnwidth]{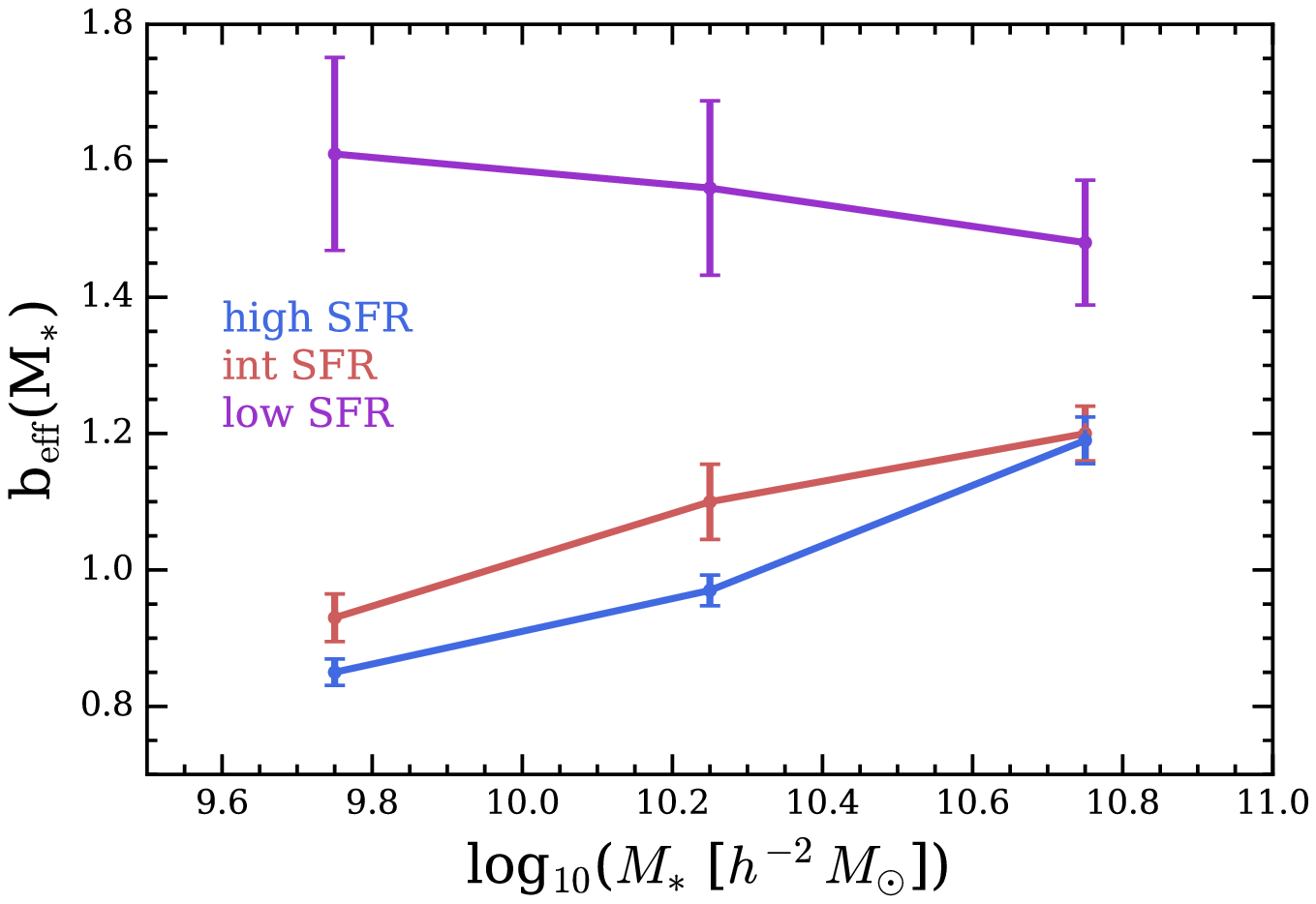}
  \caption{Effective bias, estimated using Eq.~\ref{eq:eff_bias2}, with jackknife errors for \eagle\ galaxies as a function of stellar mass, for three cuts in specific star formation rate, low (\textit{purple}), intermediate (\textit{red}) and high (\textit{blue}).
  The sample cuts are the same as in Figs.~\ref{fig:sfr-spl} and \ref{fig:sfr-histograms} and are listed in Table~\ref{tab:eagle}.}
  \label{fig:eff-bias}
\end{figure} 
 
We divide \eagle\ galaxies of a given stellar mass in three bins of star formation rate (SFR), 
the 30~per cent with the lowest star formation rate (\lq low SFR\rq), the 30~per cent with the 
highest star formation rate (\lq high SFR\rq), and the remainder (\lq int SFR\rq). Their galaxy 
clustering is plotted in Fig.~\ref{fig:sfr-spl}. The low SFR galaxies are clustered most strongly , 
which is particularly evident on the smaller scales, and this is the case in all stellar mass bins investigated. The difference in the amplitude of clustering between the high and intermediate SFR galaxies is not very large, with high SFR galaxies generally clustering least. Jackknife errors bars are not plotted to avoid clutter, but should be of order 
$\pm$0.35~dex (i.e.\ a factor $\sqrt{3}$ higher than the typical $\pm$0.2~dex errors seen in
Fig.~\ref{fig:Ms-bin-GAMA} below $r_p=1h^{-1}$~Mpc).
Any difference in clustering between the blue (\lq high SFR\rq) and red (\lq int SFR\rq) curves are therefore not very significant. At the largest scales shown, the amplitude of clustering for all populations becomes similar, with any difference now much smaller than jackknife errors. 

We do not compare these clustering measurements to those from \gama\ presented 
by Gunawardhana et al.\ ({\em in prep}), for the following main reason: the SFR and stellar mass range 
probed by \gama\ and \eagle\ are in detail poorly matched, in the sense that the limitations in 
each of the \gama\ and \eagle\ star-forming samples are hard to account for all at the same time. 
For \gama, the SFR is only measurable for galaxies with sufficiently high SFR, 
resulting in volume limited samples with SFR and stellar mass ranges restricted to  
SFR $\gtrsim 0.3 {\rm M_{\odot}\,yr^{-1}}$ and M$_* \gtrsim 10^{9.5} h^{-2}\,{\rm M}_{\odot}$.
(see Gunawardahana et al.\ ({\em in prep.})\ for details). 
Hence for a detailed comparison to take place, it would be necessary to work with samples defined by absolute cuts in SFR.
This in turn requires the \gama\ SFR measurements to be directly compatible with those in \eagle, as defining samples by SFR ranking is not possible. 
Although the galaxy number densities of \gama\ and \eagle\
split by stellar mass are in reasonably good agreement, splitting
these sub-samples by SFR does not necessarily result in samples
with similar number densities, due to differences in the bivariate
SFR-M* distribution. In fact we find that the galaxy number densities
in \eagle\ are between 60 and 70 per cent larger than in \gama\ for the same stellar mass and SFR range. This implies that a detailed
clustering comparison becomes futile: any difference observed in the clustering could be attributed to the differences in the measured
number densities of the samples.
This is in agreement with the results of \citet{Furlong2015}, who pointed out that the 
specific SFR in EAGLE are typically 0.2 to 0.5~dex lower than observed. A proper understanding of those SFR difference between data and simulations is required before a detailed and informative clustering comparison of SFR selected samples can be made.

The differences in clustering of galaxies of the same mass that are satellites versus centrals (Fig.~\ref{fig:red-blue-GAMA}), or red versus blue galaxies (Fig.~\ref{fig:sfr-spl}), should be dependent on the mass of the halo they inhabit, with more strongly clustered galaxies residing in more massive haloes. To verify that this is the case in \eagle, we plot the halo occupation distribution - the fraction of galaxies of a given $M_\star$ that inhabit haloes of mass $M_h$ - for galaxies in the stellar mass range $10^{9.5}<M_\star/h^{-2}{\rm M}_\odot<10^{10}$ split in centrals and satellites (Fig.~\ref{fig:sat-histograms}), or in bins of SFR for three ranges in $M_\star$ (Fig.~\ref{fig:sfr-histograms}).

Central galaxies in the mass range $10^{9.5}<M_\star/h^{-2}{\rm M}_\odot<10^{10}$, inhabit haloes with a narrow range of masses, from $10^{11.2} h^{-1}{\rm M}_\odot$ to $\sim 10^{13}h^{-1}{\rm M}_\odot$. In contrast, satellites of the same stellar mass, inhabit haloes 
with a wide range of masses, from $10^{11} h^{-1}{\rm M}_\odot$ to $\sim 10^{14.5}h^{-1}{\rm M}_\odot$ (Fig.~\ref{fig:sat-histograms}).
The (much) stronger clustering of satellites is therefore clearly due to the significant fraction that resides in these much more
massive (and hence more clustered) haloes \citep[see e.g.,][]{Guo2014}.

At a given stellar mass, \eagle\ galaxies with a higher star formation rate inhabit lower mass haloes than those with a lower value of SFR (at $z=0.1$), as shown in Fig.~\ref{fig:sfr-histograms} for three ranges in $M_\star$. The figure also demonstrates that the halo occupation is similar for galaxies with a high or intermediate SFR. This explains the results from Fig.~\ref{fig:sfr-spl} that, at given $M_\star$, galaxies with a low SFR cluster more strongly than those with higher SFR.

The SFR (and hence also colour) dependence of clustering in \eagle\ is related to the mechanism that causes some galaxies to have a low SFR for their mass: the reduction in SFR once a galaxy becomes a satellite\footnote{AGN feedback plays a role at higher $M_\star$ as well.}, which is discussed in detail for the \eagle\ simulations by \cite{Trayford16}. The reduction of the SFR of satellites results from two related physical processes that operate in hydrodynamical simulations: ram-pressure stripping, mainly of the outer parts of satellites as shown by \cite{Bahe15} using the {\sc gimic} simulations \citep{Crain09}}, and the strong suppression - by many orders of magnitude - of the accretion rate of gas onto satellites shown by Van de Voort et al. ({\em in prep.}) in \eagle. The much reduced gas fraction of such satellites then also implies that their interstellar medium rapidly increases in metallicity (Bah\'e et al., {\em in prep.}) - a testable prediction of the scenario.

Another way to demonstrate the bias of quenched galaxies to inhabit more massive haloes is shown in Fig.~\ref{fig:eff-bias}, which plots the effective bias of galaxies as function of stellar mass, split by SFR. The effective bias of the low SFR population is
nearly independent of $M_\star$, and considerably higher than that of the intermediate or high SFR population. For the active
galaxies with intermediate or high SFR, the bias increases with stellar mass - simply reflecting that for those galaxies SFR
increases with $M_\star$, which in turn increases with halo mass.

\section{Comparison with other models}
\label{sec:discuss}

\begin{figure}
\centering
\begin{minipage}{\columnwidth}
\includegraphics[width=\columnwidth]{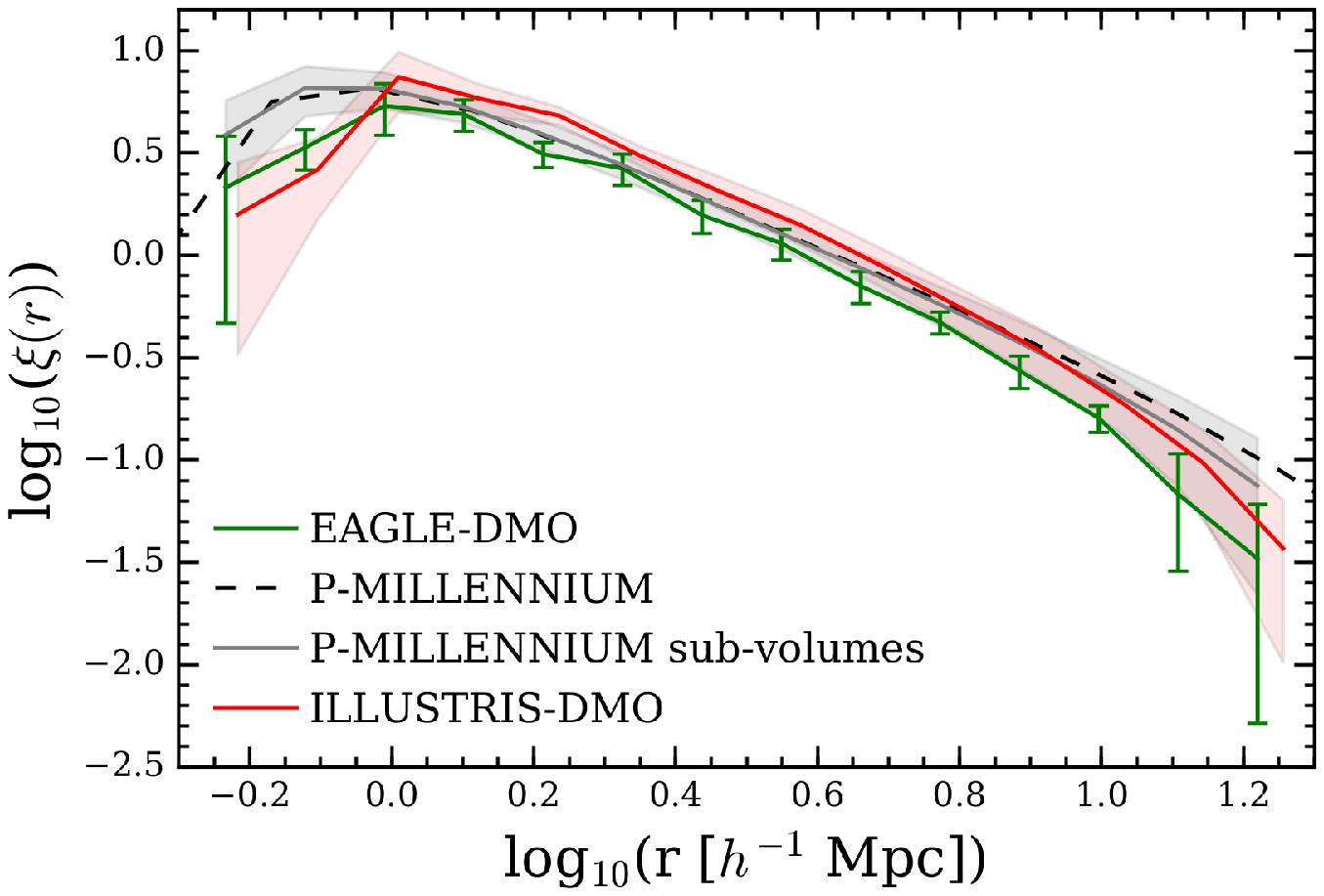}
\vspace{0.15cm}
\end{minipage}
\begin{minipage}{\columnwidth}
\includegraphics[width=\columnwidth]{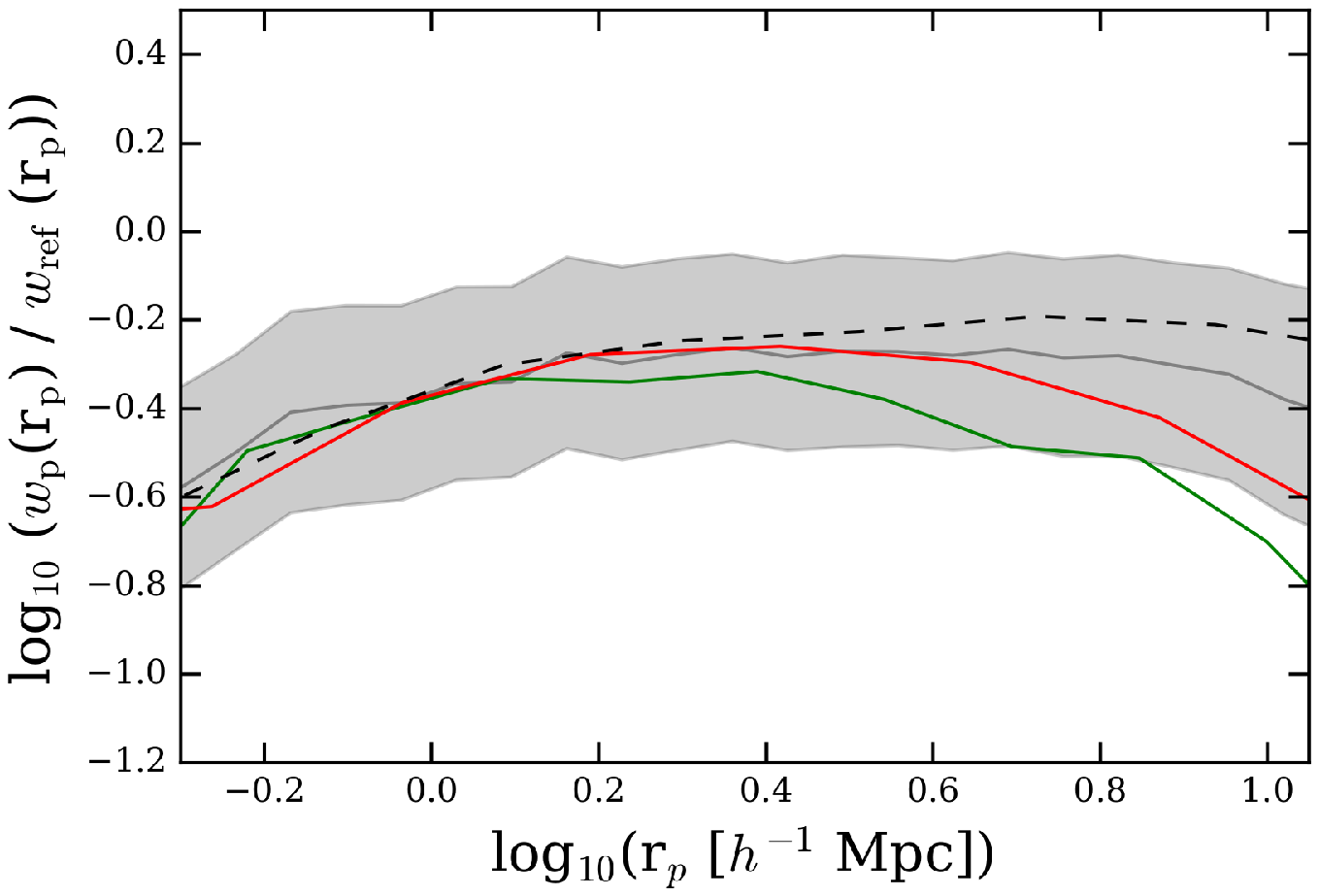}
\end{minipage}
\caption{Correlation function of $z=0$ of dark matter haloes with mass 
  $M_h> 10^{12} h^{-1} {\rm M}_{\odot}$ in real space (\textit{top panel}), and their corresponding
  projected correlation function divided by a reference power law (\textit{bottom panel}). Simulations shown are \simdmo\ (\textit{green 
  line with jackknife error bars}), \illustris\ dark matter only (\textit{red line with jackknife 
  errors represented by red shaded area}), and \pmil\ with line styles as in 
  Fig.~\ref{fig:xi_halo}.
 }
\label{fig:DMonly-comparison}
\end{figure}

\begin{figure}
  \centering
  \includegraphics[width=0.4\textwidth]{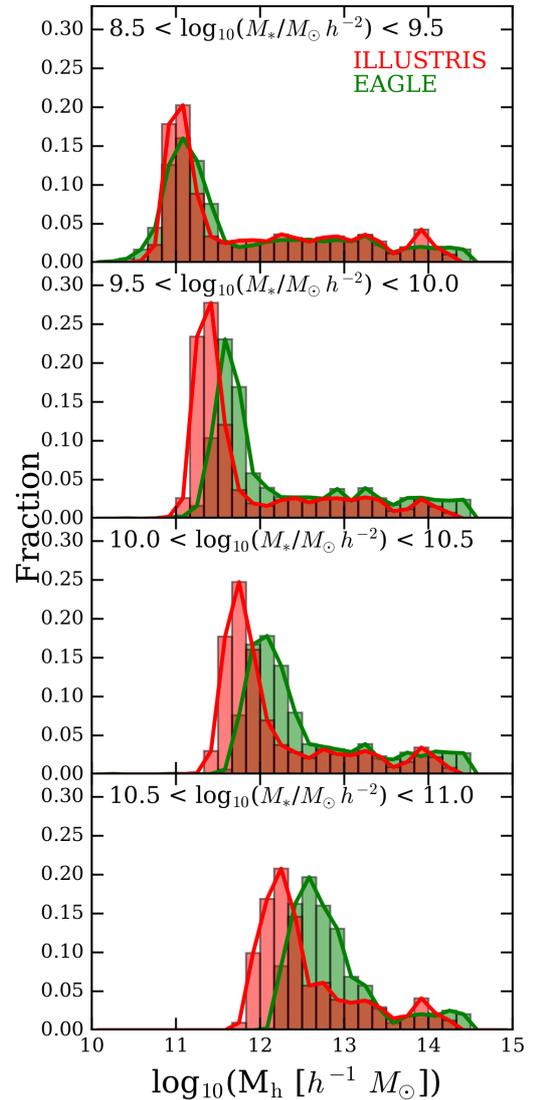}
  \caption{Halo occupation distribution of galaxies at $z=0.1$ in four different stellar mass bins, 
    as indicated in each panel, comparing \eagle\ (\textit{green}) and \illustris\ (\textit{red}). With the exception of the lowest stellar mass bin, galaxies of a given mass tend to reside in lower mass haloes in
    	\illustris\ compared to \eagle.}
  \label{fig:hist-comparison}
\end{figure}

\begin{figure}
  \centering
  \includegraphics[width=0.4\textwidth]{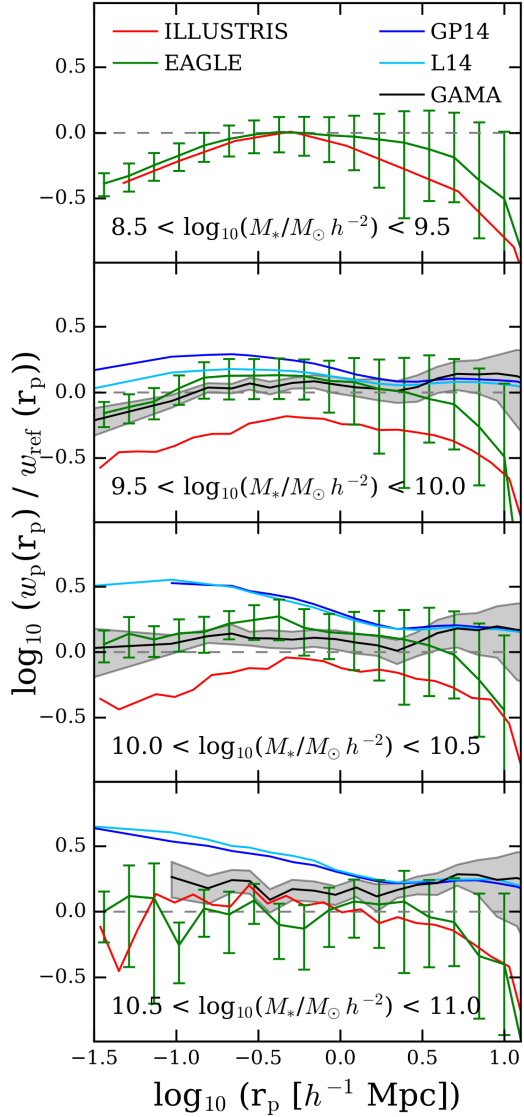}
  \caption{Comparison of galaxy clustering between different models and \gama, as a function of stellar mass ($M_\star$ increases from top to bottom).
  Different panels show the projected correlation function, $w_p(r_p)$, divided by the reference fit
  of Eq.~(\ref{eq:wp_ref}). The models shown are \eagle\ (\textit{green curve with jackknife errors})
  \illustris\ (\textit{red curve} - jackknife errors are not shown but are similar to those of \eagle), 
  GP14 (\textit{dark blue curve}) and L14 (\textit{cyan curve}). The observed correlation function from \gama\ is shown in {\textit black}
  with \textit{grey shading} encompassing jackknife errors. The latter three 
  sets of results are from \citet{Farrow15}. Models GP14 and L14
  and \gama\ curves are taken from \citealt{Farrow15}}
  \label{fig:Galform-comp-GAMA}
\end{figure}

In this section we compare the \eagle\ clustering results to two sets of models: 
{\em i)} two incarnations of the \galform\ semi-analytical model, namely the version of 
\citet{GonzalezPerez2014} (hereafter GP14), and an early version of 
\citet{Lacey2016} (hereafter L14), 
both of which were used in the \gama\ clustering study of \citet{Farrow15}; {\em ii)} the 
\illustris\ hydrodynamical simulation described by \citet{Vogelsberger2014}.

The \galform\ model assumes that galaxies form in dark matter haloes, and it uses analytical 
prescriptions to describe galaxy formation processes. These phenomenological prescriptions
have free parameters controlling different physical processes necessary for the model to be
realistic and which are tuned to fit a set of observational constraints at low redshift. 
GP14 \& L14 use halo merger trees from the {\sc millennium-MR7} simulation \citep{Guo2013}, 
which uses cosmological parameters set by {\sc wmap7} \citep{Komatsu2011}. 
The \citet{GonzalezPerez2014} model has the same physical
prescriptions as the \citet{Lagos2012} model, but set in a different cosmology to that used
by \citet{Lagos2012}, resulting in the need to retune the free parameters as described by \cite{GonzalezPerez2014}. 
The main differences between GP14 and L14 are: 
{\em i)} the assumed stellar initial mass function, with GP14 using a \cite{Kennicutt1983} 
IMF, while L14 switches from this IMF to a top-heavy IMF in star bursting galaxies 
\citep[see][for further details]{Lacey2016}; 
{\em ii)} the treatment of merging satellite galaxies, with GP14 using the Chandrasekhar 
dynamical friction timescale in an isothermal sphere as given in \cite{Lacey1993}, while 
L14 uses the \cite{Jiang2008} and \cite{Jiang2014} formula for the time-scale, which is 
empirically calibrated on N-body simulations to account for the tidal stripping of the 
accreting haloes;  {\em iii)} the assumed stellar population synthesis (SPS) model, with GP14 using an updated 
version of the \cite{Bruzual1993} SPS model, while L14 adopts the \cite{Maraston2005} SPS model.
Both GP14 and L14 were tuned to reproduced the b$_{\rm J}$-band and K-band luminosity functions of \citet{Norberg2002b}
and \citet{Cole2001}, respectively.
As \gama\ is r-band selected and as \citet{Farrow15} analysis covered a larger redshift range not
probed by those galaxy luminosity functions used to calibrated the GP14 and the L14 \galform\ models,
\citet{Farrow15} adjusted the \gama\ \galform\ lightcone mocks constructed following
\citet{Merson2013} to closely reproduce the \gama\ r-band selection function (which in turn is well described by the \gama\ r-band luminosity functions of \citet{Loveday2012,Loveday2015}.
We note that the L14 model used here and by \citet{Farrow15} has marginally different parameters from the model discussed
by \citet{Lacey2016}; we have not investigated whether this impacts any of the results presented below.

The \illustris\ simulation suite was performed with the {\sc arepo} moving-mesh code of 
\cite{Springel10}. The simulation volume, $(75 h^{-1} {\rm Mpc})^3$, is comparable to that 
of \eagle; sub-grid modules for star formation and feedback are as described by \cite{Vogelsberger2014} 
and the assumed cosmological parameters\footnote{$\Omega_{\rm m} = 0.2726$, 
$\Omega_{\Lambda} = 0.7274$, $\Omega_{\rm b} = 0.0456$, $\sigma_{8} = 0.809$, $n_{\rm s} = 0.963$, 
and H$_{0} = 100 h {\rm km s^{-1} Mpc^{-1}}$ with $h = 0.704$} 
are close to those of \cite{Planck13} assumed in \eagle. Properties of 
\illustris\ galaxies were released by the collaboration through a 
database\footnote{http://www.illustris-project.org} with content described by \cite{Nelson2015}.
The simulation reproduces several observed properties of galaxies such as for example the
colours of satellites \citep{Sales2015} and the distribution of galaxy morphologies 
\citep{Snyder2015}. However, the galaxy stellar mass function simulation has an excess 
of galaxies at both high ($M_{\star} \geqslant 10^{11.5} {\rm M}_{\odot}$) and low ($M_{\star} \leqslant 10^{10} {\rm M}_{\odot}$) 
stellar masses at redshift $z\leqslant1$, see~\cite{Vogelsberger2014}. Here we use 
the \lq Illustris-1\rq\ run (hereafter \illustris) and extract galaxy properties directly from the the \illustris\ database.

The \illustris\ simulation suite also includes dark matter-only runs. This enables us 
to compare the clustering of haloes between \simdmo\ and the dark matter-only \illustris\ (Fig.~\ref{fig:DMonly-comparison}).
The correlation function of \illustris\ 
haloes is higher than that of \simdmo, both in real and redshift space (except for the 
smallest scales plotted), with the difference consistent with sample variance as judged 
from the scatter obtained from \eagle\ like simulation sub-volumes extracted from the 
significantly larger \pmil. As discussed before, lack of large-scale power in the smaller 
boxes and the absence of integral constraint corrections on the clustering estimate cause
the correlation functions to drop below that of \pmil\ on larger scales.

Given that the \eagle\ and \illustris\ dark matter halo functions are very similar, 
whereas their galaxy stellar mass function are not \citep[see Fig.~5 of][]{Schaye15}, 
we expect some differences between the simulations in how galaxies populate the 
underlying dark matter haloes. This is indeed born-out by Fig.~\ref{fig:hist-comparison}: 
\illustris\ galaxies of given stellar mass prefer lower mass haloes, by about 0.2~dex.

In Figure~\ref{fig:Galform-comp-GAMA} we compare the two-point correlation function, $w_p$, from \eagle\ galaxies with the results from \illustris, the GP14 and L14 \galform\ models, and the \gama\ survey, in four stellar mass bins. We divided $w_p$ by the reference model $w_p^{\rm ref}$ of Eq.~(\ref{eq:wp_ref}) to decrease the dynamic range in the plot.
The shape of the correlation functions of \eagle\ and \illustris\ are very similar (we do not plot jackknife errors on
the \illustris\ curves to avoid clutter, but they are nearly identical to those of \eagle), 
but with \illustris\ offset to
smaller values except for the lowest bin in stellar mass (the top panel). The poor sampling of large-scale modes in
both hydrodynamical simulations, combined with the integral constraint, may lead to an net offset of $w_p$ - as we
demonstrated explicitly in Fig.~\ref{fig:xi_halo} for the dark matter haloes. The level of the offset is consistent with
sample variance - as shown by the comparing to the \pmil\ results. However, somewhat surprisingly, whereas haloes in
\illustris\ are more strongly clustered than those in \eagle\ (red line above green line in Fig.~\ref{fig:DMonly-comparison}), 
\illustris\ galaxies are {\em less} strongly clustered than those in \eagle\ at given $M_\star$ 
(Fig.~\ref{fig:Galform-comp-GAMA}). This is related to the differences in stellar mass function: the galaxy number
density is higher in \illustris\ compared to \eagle, therefore \illustris\ galaxies of given $M_\star$ inhabit haloes of
lower mass (Fig.\ref{fig:hist-comparison}) which are less clustered. This effect is relatively small, however, 
and we conclude that the clustering is consistent in both models given the relatively large jackknife errors.

\begin{figure}
  \centering
  \includegraphics[width=\columnwidth]{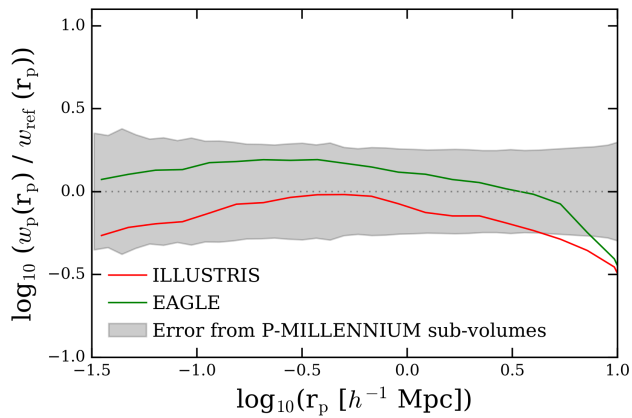}
  \caption{Clustering of the most massive galaxies corresponding to a mean number density of $n_{\rm gal} \sim 1.49 10^{-2} h^3 {\rm Mpc}^{-3}$ at $z=0.1$. The projected correlation function $w_p$ is divided by the reference model $w_p^{\rm ref}$ of Eq.~(\ref{eq:wp_ref}). The \eagle\ and \illustris\ simulations are plotted in \textit{green} and \textit{red}, respectively. Sample variance in the clustering of dark matter haloes, mass ranked and selected to have the same mean number density, estimated using from sub-sampling the \pmil\ 
  volume, are shown by the \textit{grey shaded area} (computed with Eq.~\ref{eq_PM}, and including the effective bias of the sample).}
  \label{fig:Comparison-eqDensity}
\end{figure}

We can partially compensate for differences in the stellar mass function of \eagle\ and \illustris\ by comparing galaxy clustering at a given number density, rather than stellar mass. To do so, we
select all \eagle\ galaxies with $M_{\star} > 10^{9.5}~h^{-2}~{\rm M}_{\sun}$, yielding a galaxy number density of $n_{\rm gal}\simeq 1.49\times 10^{-2} h^3 {\rm Mpc}^{-3}$ and then find the corresponding stellar mass 
range of $M_{\star} > 10^{9.73}~h^{-2}~{\rm M}_{\sun}$ in \illustris\ above which the number density is equal to $n_{\rm gal}$. The clustering of these two samples is compared in Figure~\ref{fig:Comparison-eqDensity}, with \illustris\ in red and 
\eagle\ in green, with sample variance in the clustering of dark matter haloes with the same number density estimated by sub-sampling \pmil\ volumes in grey. Selected in this way, the clustering in \eagle\ is higher than in \illustris, but the differences are consistent with sample variance.
	
We now turn to the semi-analytical models plotted in Fig.~\ref{fig:Galform-comp-GAMA}. 
The \galform\ model shown are the average of 26 \gama\ light cone mocks as described in \citet{Farrow15}.
Like in that paper, we assume that the errors on those model clustering
results are negligible compared to the errors measured on the \gama\ sample. See \citet{Farrow15}
for a quantitative description of how adequate the GP14 and L14 \galform\ models are in describing the observed \gama\ clustering.
The correlation functions of GP14 and L14 are very similar, except in the second panel from the
top where the GP14 model is above that of L14 on scales below $r_p\sim 1h^{-1}$~Mpc.
Both models show stronger clustering than observed below $r_p\sim 1h^{-1}$~Mpc. As discussed by \cite{Farrow15}, the high values of $w_p$
in the \galform\ models are caused by an excess of satellites galaxies and/or their radial distribution in clusters. The
satellite merging scheme is the principal mechanism which impacts directly on the number of satellites within haloes. The two versions of \galform\ we use, include the default scheme in which satellite galaxies merge onto the central galaxy after an analytically determined dynamical friction merger time-scale. \citet{Campbell2015} compared the standard \galform\ scheme with a different one \citep{Simha16}, in which the merger time-scale make use of the information from dark matter subhalo of each satellite galaxy. They find that the new scheme reduces the amplitude at small scales and show good agreement with observational results.  McCullagh et al. ({\em prep})
and \citet{Gonzalez2017} show that implementing the \citet{Simha16} merger scheme within the GP14 model and applying it to the \pmil\ simulation results in clustering measurements that are in significantly better agreement with the observed ones on small scales. More detailed studies of the \galform\ clustering predictions on small scales are needed to address the known limitations of samples split by colour \citep[see e.g. Fig. 14 of][]{Farrow15}.
At large-scales the semi-analytic models show a good agreement with observational data.

To summarise: we find that both \eagle\ and \illustris\ reproduce the clustering of galaxies in \gama\ on scales below
$r_p\sim 4h^{-1}$~Mpc, but both jackknife errors and sample variance are still relatively large. Above this scale, both simulations are affected by the relatively small simulation volume. The good agreement show that these models both reproduce the dominant effects of environment on satellites, a crucial ingredient in getting $w_p$ right. Larger simulation volumes, that would yield smaller errors, are needed to make the clustering constraint more stringent. The hydrodynamical models perform better on these smaller scales than the \galform\ models of GP14 and L14.

\section{Conclusions}
\label{sect:conclusions}

We have studied the two-point correlation function of galaxies at $z=0.1$ in the \eagle\ cosmological hydrodynamical simulation \citep{Schaye15}. The sub-grid parameters of \eagle\ are calibrated as described by \cite{Crain15}
to the present-day galaxy stellar mass function (amongst other observables, such as galaxy sizes), but the clustering
properties of galaxies is a prediction of the simulation. We have compared the results to the clustering of observed
galaxies from the \gama\ survey \citep{Driver2011}, as well as two incarnations of the \galform\ semi-analytical model
(the GP14 model described by \citet{GonzalezPerez2014}, and an early version of the
 \citet{Lacey2016} model, referred to as L14 model described in \citet{Farrow15}), and the \illustris\ simulation described by \cite{Vogelsberger2014}.

The simulation volume of the largest \eagle\ simulation we use here, is still relatively small at ($100~{\rm Mpc}$)$^3$. We examine how lack of (and poor sampling of) large-scale modes and sample variance might affect clustering by comparing the real-space clustering of dark matter haloes with $M_{h} > 10^{12} h^{-1}\,{\rm M}_{\odot}$ in a dark matter-only version of \eagle\ (called \simdmo) to that in the much larger volume, ($800~{\rm Mpc}$)$^3$, of the \pmil\ dark matter only simulation that uses the same \cite{Planck13} cosmology. We find that the clustering amplitude is similar in the range $r \sim 1 - 6 \, h^{-1}\,{\rm Mpc}$, while the clustering amplitude in \simdmo\ is smaller 
on larger scales. We therefore focus our attention on scales up to $\sim 6\,h^{-1}$~Mpc. 	We also show that jackknife-error estimates of the clustering amplitude calculated for \simdmo\ are similar to the variance measured between \eagle-sized volumes drawn from \pmil. This encourages us to quote jackknife errors for clustering of \eagle\ galaxies as well.

We use the optical broad-band magnitudes and colours of \eagle\ galaxies calculated using the fiducial model of \cite{Trayford15}.
Briefly, this calculation combines the \cite{bc03} population synthesis model for stars with a two-component screen model for dust. The dust model is based on that of \cite{Charlot2000}, with dust-screen optical depth depending on the enriched star-forming gas content of galaxies, and including additional scatter to represent orientation effects.

\cite{Trayford15} show that the r-band luminosity function of EAGLE agrees well with observations. The luminosity-dependent fraction of red and blue galaxies also performs reasonably well, although an excess of blue galaxies is found at high mass. Using these predicted fluxes allows us to compared the projected two-point correlation function, $w_p(r_p)$ (defined in Eq.~\ref{eq:wp_rp}) in \eagle\ directly to that measured in \gama\, for galaxies selected by r-band absolute magnitude $M_{r,h}$ and/or $(g-r)_{0}$ rest-frame colour, in addition to stellar mass. We do so using the (nearly) volume-limited sample of \gama\ galaxies described by \cite{Farrow15}.

Our findings can be summarised as follows:
\begin{itemize}
 \item The (projected) clustering of \eagle\ galaxies in bins of stellar mass agrees well with that from \gama, with differences consistent within the errors, and we find similar good agreement for galaxies selected in bins of $r$-band absolute magnitude, $M_{r,h}$. Given that the number densities of these galaxies in simulation and data agree as well, this gives us confidence that \eagle\ galaxies of given mass or luminosity, inhabit haloes of similar mass as those in \gama. The observed clustering amplitude increases with mass and luminosity. This trend is not seen in \eagle, however the \eagle\ clustering results are still consistent with the data given the relatively large jackknife errors and the effect of missing large-scale power.
\item At a given stellar mass, red \eagle\ galaxies are more strongly clustered than blue galaxies. In \eagle, red galaxies are either satellites - with ram-pressure stripping of gas \citep{Bahe15} and a reduction in the cosmological accretion rate of fresh gas (Van de Voort et al., {\em in prep.}) both playing in role in reducing the star formation rate and making the galaxy red, or their star formation is reduced by their central black hole \citep{Trayford16}. The stronger clustering of red galaxies is then a consequence of the higher halo mass they inhabit, compared to blue galaxies of the same $M_\star$.  The difference in clustering amplitude between red and blue galaxies is too strong in \eagle\ compared to \gama. This overabundance of red galaxies in \eagle\ is at least partly due to lack of numerical resolution \citep{Schaye15,Trayford16}, although poor sampling of massive groups and clusters plays a role as well.
\item On small scales, low SFR galaxies cluster more strongly for all the stellar mass bins studied. This is because galaxies with low SFR at a given mass tend to be satellites, and as before, the enhanced clustering reflects that of their more massive dark matter hosts.
 \end{itemize}

We conclude that the galaxy clustering predicted by \eagle\ is in very good agreement with \gama\ on projected scales up to $r_p\sim 4h^{-1}$~Mpc, also when galaxies are split by colour. \eagle\ and observed galaxies therefore inhabit haloes of similar mass, and the reduction in the SFR of \eagle\ galaxies when they become a satellite, mimics that of observed galaxies. However, the limited simulation volume of the simulation yields relatively large jackknife errors as well as large sample variance. Better tests of the realism of \eagle\ requires clustering studies in somewhat large volumes.

Comparing to other models, we find that both the GP14 and L14 semi-analytical models overestimate the galaxy clustering amplitude at small scales, $r_p\lesssim 1h^{-1}$~Mpc, while showing good agreement with \gama\ on larger scales. We speculate that the excess at small scales is caused by the satellite merging schemes implemented, which are crucial and impact directly on the number of satellites and their radial distribution \citep{Contreras2013,Campbell2015}.
The \illustris\ simulation yields very similar clustering measures to \eagle. At a given stellar mass, the clustering
amplitude in \illustris\ is lower than in \eagle, although the difference is consistent given the jackknife error
estimates. This good agreement is slightly fortuitous: the fact that \illustris\ galaxies tend to inhabit haloes of
lower mass than \eagle\ galaxies
\citep[consistent with \illustris\ over predicting the galaxy stellar mass function for most
values of $M_\star$,][- which would yield lower clustering - is partially compensated by
\illustris\ haloes clustering more strongly than \eagle\ haloes (with the difference consistent with sample variance]{Vogelsberger2014}

Galaxy clustering measurements provide powerful constraints on galaxy formation models. Here we have shown that the \eagle\ simulation reproduces the spatial distribution of galaxies measured in the \gama\ survey, even when galaxies are split by stellar mass, luminosity and colour. This increases our confidence in the realism of the simulation. However, sample variance is still relatively large, given the small volume simulated, and better constraints require larger simulations, even when studying clustering on the smaller scales where the galaxy formation modelling is tested most
	stringently.
	
\section*{Acknowledgement}
The authors would like to thank the referee, Manodeep Sinha, and Joop Schaye and Robert Crain
for useful discussions.
MCA acknowledges support from the European Commission's Framework Programme 7, through the Marie Curie International
Research Staff Exchange Scheme LACEGAL (PIRSES--GA--2010--269264). This work has been partially supported by PICT Raices 2011/959 of Ministery of Science (Argentina). IZ is supported by NSF grant AST-1612085 and by a CWRU Faculty Seed Grant. MCA, SP and IZ acknowledge the hospitality of the ICC at Durham, where this project was started.
This work was supported by the Science and Technology Facilities Council [grant number ST/F001166/1], by the Interuniversity Attraction Poles Programme initiated by the Belgian Science Policy Office ([AP P7/08 CHARM]). We used the DiRAC Data Centric system at Durham University, operated by the Institute for Computational Cosmology on behalf of the STFC DiRAC HPC Facility (www.dirac.ac.uk). This equipment was funded by BIS National E-Infrastructure capital grant ST/K00042X/1, STFC capital grant ST/H008519/1, and STFC DiRAC is part of the National E-Infrastructure. 
PN acknowledges the support of the Royal Society through the award of a University Research Fellowship and the European Research
Council, through receipt of a Starting Grant (DEGAS-259586).
TT, PN, RGB and MS acknowledge the support of the Science and Technology Facilities Council (ST/L00075X/1 \& ST/P000541/1).
The data used in the work is publically available in the \eagle\ database described by \cite{McAlpine16}.
\gama\ is a joint European-Australasian project based around a spectroscopic
campaign using the Anglo-Australian Telescope. The \gama\ input catalogue is
based on data taken from the Sloan Digital Sky Survey and the UKIRT Infrared
Deep Sky Survey. Complementary imaging of the \gama\ regions is being obtained
by a number of independent survey programmes including GALEX MIS, VST KiDS,
VISTA VIKING, WISE, Herschel-ATLAS, GMRT and ASKAP providing UV to
radio coverage. \gama\ is funded by the STFC (UK), the ARC (Australia), the AAO,
and the participating institutions. The \gama\ website is http://www.gama-survey.org/ .
\bibliographystyle{mnras}

\bibliography{Clustering}

\IfFileExists{\jobname.bbl}{}
{\typeout{}
\typeout{****************************************************}
\typeout{****************************************************}
\typeout{** Please run "bibtex \jobname" to optain}
\typeout{** the bibliography and then re-run LaTeX}
\typeout{** twice to fix the references!}
\typeout{****************************************************}
\typeout{****************************************************}
\typeout{}
}
\end{document}